\documentstyle[aps]{revtex}

\begin{document}
\draft
\title{Dynamical Decimation Renormalization-Group Technique : Kinetic Gaussian
Model on Non-Branching, Braching and Multi-branching Koch Curve}
\author{Jian-Yang Zhu$^{1,2,3}$\thanks{%
Electronic address: zhujy@bnu.edu.cn} and Z. R. Yang$^{1,2}$}
\address{$^1$CCAST (World Laboratory), Box 8730, Beijing 100080, China\\
$^2$Department of Physics and Institute of Theoretical Physics,Beijing
Normal University, Beijing 100875, China\thanks{%
Mailing address.}\\
$^3$Department of Physics, Jiangxi Normal University, Nanchang 330027, China }
\maketitle

\begin{abstract}
A generalizing formulation of dynamical real-space renormalization that
suits for arbitrary spin systems is suggested. The new version replaces the
single-spin flipping Glauber dynamics with the single-spin transition
dynamics. As an application, in this paper we mainly investigate the
critical slowing down of the Gaussian spin model on three fractal lattices,
including nonbranching, branching and multibranching Koch Curve. The
dynamical critical exponent $z$ is calculated for these lattices using an
exact decimation renormalization transformation in the assumption of the
magnetic-like perturbation, and a universal result $z=1/\nu $ is found.
\end{abstract}

\pacs{PACS numbers: 64.60.Ht, 64.60.Ak, 05.50.+q, 75.10.Hk}

\section{Introduction}

The dynamics of spin systems approaching their second-order phase transition
points have been an important subject of many studies in the last few
decades. One of the interesting phenomena is the critical slowing down
characterized by a divergent relaxation time $\tau $. A reasonable
explanation is seemingly that the long range fluctuation leads to long time
evolution of the order parameter. According to the dynamical scaling
hypothesis\cite{Halperin-1}, the divergent relaxation time $\tau $ and the
divergent correlation length $\zeta $ can be related by $\tau \sim \zeta ^z,$
where $z$ is the dynamical critical exponent and is believed to depend only
on large universal features of the model Hamiltonian and the assumed dynamic
process\cite{Halperin-2}.

Obtaining the exact solution based on a master equation, except for a few
cases\cite{Glauber,Zhu-1,Zhu-2}, is not an easy job. One have to evaluate it
by means of approximate methods, such as the Monte Carlo simulation, the
high-temperature series expansion, etc. However, the success of
renormalization-group (RG) methods\cite{Wilson,Ma} in obtaining the critical
exponents and universality classes of static problems led to several
attempts to use RG ideas in critical dynamics. One of the typical examples%
\cite{Halperin-2} is a generalization of the $\varepsilon $ expansion
technique in which the dynamics is described by a Langevin equation. This
method enables the calculation of the time-dependent correlation functions,
but it is only useful near the upper critical dimension. Another\cite{Achiam}
is a generalization of the real-space RG techniques, the starting point is a
master equation instead of Hamiltonian. This method is preferable in
discrete spin systems, and could be used to directly calculate the dynamical
critical exponent. In addition, it is simple and transparent, and very
accurate for certain systems so that it has been used quite extensively in
the last years \cite{1,2,3,4,5,6,7}. Other examples see Refs.\cite
{Mazenko,Kinzel}.

The dynamical real-space renormalization group (DRSRG) technique, proposed
by Y. Achiam and J. M. Kosterlitz\cite{Achiam} and perfected by D.Kandel\cite
{Kandel}, is our focus in this paper. First, we establish a formulation of
DRSRG applying to arbitrary spin systems. Then, we investigate the critical
slowing down of the continuous spin model on different fractal lattices. In
the generalizing formulation of DRSRG, we replace the single-spin flipping
Glauber dynamics\cite{Glauber} with the single-spin transition dynamics\cite
{Zhu-1}, and use the same notation of Ref.\cite{Kandel} to express the
critical dynamical exponent $z.$

During the last many years, scientific journals have published many papers
concerning critical dynamics of discrete spin systems, but a systematic
study of the critical dynamics in continuous spin systems is very lacked
indeed. The purpose of our latest papers\cite{Zhu-1,Zhu-2} and this work is
a attempt filling this gap. This is just our main motivation. We realize the
fact that, though the Gaussian model is certainly an idealization, it is
interesting and simple enough to obtain some fundamental knowledge of
dynamical process in cooperative systems. So this is a ideal dynamical model
that interests us greatly. We also realize that, as an extension of Ising
model, the Gaussian model shows much differences from Ising model in the
properties of static phase transition, and yet its knowledge of the
dynamical behavior is unclear. Within the framework of single-spin
transition critical dynamics in our previous paper\cite{Zhu-1}, we have
obtained the dynamical critical exponent of the Gaussian model, $z=1/\nu =2$%
, at the critical point $K_c=b/2d$ based on rigorous calculation. This means
that the dynamical exponent is highly universal on translational symmetric
lattices. However, what is the dynamical exponent on dilational symmetric
lattice systems? All of these motivate us to finish this work.

This paper is organized as follows: Section \ref{Sec.2} is a detailed
description of the dynamical real-space renormalization group (DRSRG)
technique in which the dynamics is described by a Markov process with the
single-spin transition instead of the single-spin flipping. In Sec. \ref
{Sec.3}, the critical dynamics of Gaussian model on three fractal lattices
is studied. We take the exact decimation transformation and calculate the
dynamical critical exponent $z$ in the assumption of the magnetic-like
perturbation$.$ Section \ref{Sec.4} is our summary and discussion.

\section{Description of the method}

\label{Sec.2}

In the single-spin transition critical dynamics\cite{Zhu-1}, the master
equation can be written as 
\begin{equation}
\frac d{dt}P\left( \left\{ \sigma \right\} ,t\right) =-\sum_i\sum_{\hat{%
\sigma}}\left( 1-\hat{p}_i\right) W_i\left( \sigma _i\rightarrow \hat{\sigma}%
_i\right) P\left( \left\{ \sigma \right\} ,t\right) ,  \eqnum{2.1}
\label{ME}
\end{equation}
where $p_i$ is the transition operation defined by 
\[
p_if\left( \sigma _1,\sigma _2,\ldots ,\sigma _i,\ldots ,\sigma _N,t\right)
=f\left( \sigma _1,\sigma _2,\ldots ,\hat{\sigma}_i,\ldots ,\sigma
_N,t\right) , 
\]
and $W_i\left( \sigma _i\rightarrow \hat{\sigma}_i\right) $ is the
single-spin transition probability which satisfies the following restraint
conditions:

(a) Ergodicity: 
\begin{equation}
\forall \sigma _j,\hat{\sigma}_j:\ W_j(\sigma _j\rightarrow \hat{\sigma}%
_j)\neq 0;  \eqnum{2.2(a)}  \label{2a}
\end{equation}

(b) Positivity: 
\begin{equation}
\forall \sigma _j,\hat{\sigma}_j:\ W_j(\sigma _j\rightarrow \hat{\sigma}%
_j)\geq 0;  \eqnum{2.2(b)}  \label{2b}
\end{equation}

(c) Normalization: 
\begin{equation}
\forall \sigma _j:\ \sum_{\hat{\sigma}_j}W_j(\sigma _j\rightarrow \hat{\sigma%
}_j)=1;  \eqnum{2.2(c)}  \label{2c}
\end{equation}

(d) Detailed balance: 
\begin{equation}
\forall \sigma _j,\hat{\sigma}_j:\ \frac{W_j(\sigma _j\rightarrow \hat{\sigma%
}_j)}{W_j(\hat{\sigma}_j\rightarrow \sigma _j)}=\frac{P_{{\normalsize eq}%
}(\sigma _1,\ldots ,\hat{\sigma}_j,\ldots ,\sigma _N)}{P_{{\normalsize eq}%
}(\sigma _1,\ldots ,\sigma _j,\ldots ,\sigma _N)},  \eqnum{2.2(d)}
\label{2d}
\end{equation}
in which 
\[
P_{{\normalsize eq}}\left( \{\sigma \}\right) =\frac 1Z\exp [-\beta {\cal H}%
(\{\sigma \})],\quad Z=\sum_{\{\sigma \}}\exp [-\beta {\cal H}(\{\sigma
\})], 
\]
where $P_{eq}$ is the equilibrium Boltzmann distribution function, $Z$ the
partition function and ${\cal H}(\{\sigma \})$ the system Hamiltonian. A
well-chosen form of the transition probability is 
\begin{equation}
W_i(\sigma _i\rightarrow \hat{\sigma}_i)=\frac 1{Q_i}\exp \left[ -\beta 
{\cal H}_i\left( \hat{\sigma}_i,\sum_{\left\langle {i,j}\right\rangle
}\sigma _j\right) \right] =\frac{\exp \left[ -\beta {\cal H}_i\left( \hat{%
\sigma}_i,\sum_{\left\langle {i,j}\right\rangle }\sigma _j\right) \right] }{%
\sum_{\hat{\sigma}_i}\exp \left[ -\beta {\cal H}_i\left( \hat{\sigma}%
_i,\sum_{\left\langle {i,j}\right\rangle }\sigma _j\right) \right] }. 
\eqnum{2.3}  \label{W}
\end{equation}

In order to study the critical slowing down, we can limit ourselves to the
relaxation of an infinitely small perturbation from equilibrium. Following
Y. Achiam's idea\cite{4}, two selections can be considered, that is
magnetic-like perturbation 
\begin{equation}
P\left( \left\{ \sigma \right\} ,t\right) =\left[ 1+\sum_ih_{q_i}\left(
t\right) \sigma _i\right] P_{{\normalsize eq}}\left( \left\{ \sigma \right\}
\right)  \eqnum{2.4}  \label{magnetic-like}
\end{equation}
or energy-like perturbation 
\begin{equation}
P\left( \left\{ \sigma \right\} ,t\right) =\left[ 1+\sum_{\left\langle {i,j}%
\right\rangle }h_{q_i}^E\left( t\right) \sigma _i\sigma _j\right] P_{%
{\normalsize eq}}\left( \left\{ \sigma \right\} \right) ,  \eqnum{2.5}
\label{energy-like}
\end{equation}
where $q_i$ distinguishes between points which have different $R$ (the order
of ramification), $\left\langle i,j\right\rangle $ denotes a sum over
nearest-neighbor pairs, and $h_{q_i}$ and $h_{q_i}^E$ are the reduced
external fields.

Based on these two considerations (\ref{magnetic-like}) or (\ref{energy-like}%
), the master equation (\ref{ME}) takes the following forms 
\begin{equation}
\frac d{dt}\sum_ih_{q_i}\left( t\right) \sigma _iP_{{\normalsize eq}}\left(
\left\{ \sigma \right\} \right) =-\sum_i\sum_{\hat{\sigma}_i}h_{q_i}\left(
t\right) \left( \sigma _i-\hat{\sigma}_i\right) W_i\left( \sigma
_i\rightarrow \hat{\sigma}_i\right) P_{{\normalsize eq}}\left( \left\{
\sigma \right\} \right) ,  \eqnum{2.6}  \label{ME1}
\end{equation}
or 
\begin{equation}
\frac d{dt}\sum_{\left\langle {i,j}\right\rangle }h_{q_i}^E\left( t\right)
\sigma _i\sigma _jP_{{\normalsize eq}}\left( \left\{ \sigma \right\} \right)
=-\sum_{\left\langle {i,j}\right\rangle }\sum_{\hat{\sigma}}h_{q_i}^E\left(
t\right) \left( \sigma _i-\hat{\sigma}_i\right) \sigma _jW_i\left( \sigma
_i\rightarrow \hat{\sigma}_i\right) P_{{\normalsize eq}}\left( \left\{
\sigma \right\} \right) ,  \eqnum{2.7}  \label{ME2}
\end{equation}
respectively. We can express further the Eqs. (\ref{ME1}) and (\ref{ME2}) as
the unitized matrix formulation 
\begin{equation}
\frac d{dt}{\bf h}\left( t\right) {\bf \cdot \Lambda }\left( \sigma \right)
P_{{\normalsize eq}}\left( k,\left\{ \sigma \right\} \right) =-{\bf h}\left(
t\right) {\bf \cdot \Omega }\left( k,\sigma \right) P_{{\normalsize eq}%
}\left( k,\left\{ \sigma \right\} \right) ,  \eqnum{2.8}  \label{eq}
\end{equation}
where, ${\bf h}\left( t\right) $ is a row matrix, ${\bf \Lambda }\left(
\sigma \right) $ and ${\bf \Omega }\left( k,\sigma \right) $ are column
matrices.

The critical dynamical behavior of the system described by Eq. (\ref{eq}),
can be studied using the dynamical real-space renormalization group (DRSRG)
technique. The DRSRG is composed of two stages. The first stage is the
rescaling of the space by 
\begin{equation}
x\rightarrow x^{\prime }=Lx.  \eqnum{2.9}  \label{space}
\end{equation}
which is performed using a RG transformation (such as decimation or
site-block transformation), where $L$ is the length-rescaling factor. For
example, in the case of the decimation transformation, the spins are divided
into two groups $\left\{ \sigma \right\} $ and $\left\{ \mu \right\} $ under
the control of a decimation operator $T\left( \mu ,\sigma \right) $, then a
trace over the $\left\{ \sigma \right\} $ is performed. The process of
decimation for a spin function $f\left( \left\{ \sigma \right\} \right) $
can be demonstrated as

\begin{equation}
R\left\{ f\left( \left\{ \sigma \right\} \right) \right\} =\sum_{\left\{
\sigma \right\} }T\left( \mu ,\sigma \right) f\left( \left\{ \sigma \right\}
\right) =f\left( \left\{ \mu \right\} \right) .  \eqnum{2.10}  \label{R}
\end{equation}
It is certain that we need to rescale the interaction parameter $k=J/k_\beta
T$, and the spin $\mu $, i.e. 
\begin{equation}
k\rightarrow k^{\prime }=R\left( k\right) k,\ \ \mu \rightarrow \mu ^{\prime
}=\xi \left( k\right) \mu ,  \eqnum{2.11}  \label{rescale}
\end{equation}
so as to keep on an invariant form of the probability distribution, $%
P_{eq}^{\prime }\left( k^{\prime },\left\{ \mu ^{\prime }\right\} \right) $,
where $\xi \left( k\right) $ is the spin-rescaling factor.

With the decimation transformation (\ref{R}), Eq. (\ref{eq}) takes the form 
\begin{equation}
\frac d{dt}{\bf h}\left( t\right) {\bf \cdot R_\Lambda \left( k\right) \cdot
\Lambda }^{\prime }\left( \mu ^{\prime },k^{\prime }\right) P_{{\normalsize %
eq}}^{\prime }\left( k^{\prime },\left\{ \mu ^{\prime }\right\} \right) =-%
{\bf h}\left( t\right) {\bf \cdot R_\Omega \left( k\right) \cdot \Omega }%
^{\prime }\left( \mu ^{\prime },k^{\prime }\right) P_{{\normalsize eq}%
}^{\prime }\left( k^{\prime },\left\{ \mu ^{\prime }\right\} \right) , 
\eqnum{2.12}  \label{2.12}
\end{equation}
where ${\bf \Lambda }^{\prime }\left( \mu ^{\prime },k^{\prime }\right) $
and ${\bf \Omega }^{\prime }\left( \mu ^{\prime },k^{\prime }\right) $
retain the original form. Taking the monomark 
\[
{\bf h}^{\prime }\left( t,k^{\prime }\right) ={\bf h}\left( t\right) {\bf %
\cdot R_\Lambda \left( k\right) ,} 
\]
which can be regarded as a RG transformation of the dynamic parameter ${\bf h%
}\left( t\right) $, Eq. (\ref{2.12}) can be rewritten as 
\begin{eqnarray}
&&\frac d{dt}{\bf h}^{\prime }\left( t,k^{\prime }\right) {\bf \cdot \Lambda 
}^{\prime }\left( \mu ^{\prime },k^{\prime }\right) P_{{\normalsize eq}%
}^{\prime }\left( k^{\prime },\left\{ \mu ^{\prime }\right\} \right) 
\nonumber \\
&=&-{\bf h}^{\prime }\left( t,k^{\prime }\right) {\bf \cdot }\left[ {\bf %
R_\Lambda ^{-1}\left( k\right) \cdot R_\Omega \left( k\right) }\right] {\bf %
\cdot \Omega }^{\prime }\left( \mu ^{\prime },k^{\prime }\right) P_{%
{\normalsize eq}}^{\prime }\left( k^{\prime },\left\{ \mu ^{\prime }\right\}
\right) .  \eqnum{2.13}  \label{2.13}
\end{eqnarray}

The second stage of the DRSRG is the rescaling of the time by 
\begin{equation}
t\rightarrow t^{\prime }=L^{-z}t,  \eqnum{2.14}  \label{time}
\end{equation}
which should be resulted in that Eq. (\ref{2.12}) is restored to the
invariant form of the master equation (\ref{eq}):

\begin{equation}
\frac d{dt^{\prime }}{\bf h}^{\prime }\left( t^{\prime },k^{\prime }\right) 
{\bf \cdot \Lambda }^{\prime }\left( \mu ^{\prime },k^{\prime }\right) P_{%
{\normalsize eq}}^{\prime }\left( k^{\prime },\left\{ \mu ^{\prime }\right\}
\right) =-{\bf h}^{\prime }\left( t^{\prime },k^{\prime }\right) {\bf \cdot
\Omega }^{\prime }\left( \mu ^{\prime },k^{\prime }\right) P_{{\normalsize eq%
}}^{\prime }\left( k^{\prime },\left\{ \mu ^{\prime }\right\} \right) . 
\eqnum{2.15}  \label{eq*}
\end{equation}

We might encounter two different cases in carrying out the Eq. (\ref{eq*}).
It can be realized via the following analyses.

First, for some homogeneous lattices with same coordination number ($q_i=q,$ 
$h_{q_i}=h$), ${\bf R}_{{\bf \Lambda }}\left( k\right) $ and ${\bf R}_{{\bf %
\Omega }}\left( k\right) $ are only $1\times 1$ matrices. When system
approaches its critical point $k_c$, $\lambda \equiv {\bf R}_{{\bf \Lambda }%
}\left( k_c\right) =$ const, $\omega \equiv {\bf R}_{{\bf \Omega }}\left(
k_c\right) =$ const, then from Eq. (\ref{2.13}) we can see that the
invariant form of the master equation (\ref{eq}) can be restored by
preforming the time rescaling 
\begin{equation}
t\rightarrow t^{\prime }=L^{-z}t=\frac t{\lambda /\omega },  \eqnum{2.16}
\label{2.16}
\end{equation}
and from here we can further obtain the dynamical critical exponent $z$ 
\begin{equation}
z=\frac{\ln \left( \lambda /\omega \right) }{\ln L}.  \eqnum{2.17}  \label{z}
\end{equation}

Second, for some inhomogeneous lattices with different coordination number, $%
{\bf R}_{{\bf \Lambda }}\left( k\right) $ and ${\bf R}_{{\bf \Omega }}\left(
k\right) $ are $m\times m$ square matrices in which the order $m$ of the
matrices depends on the number of the parameter $h_{q_i}$. In this case we
have to look for the invariant form at the limit of the order $n\rightarrow
\infty $ of the RG transformation. Because our starting point is very close
to the fixed point $k_c$ of the static RG transformation, the eigenvalues of
the transformation matrices ${\bf R}_{{\bf \Lambda }}\left( k\rightarrow
k_c\right) $ and ${\bf R}_{{\bf \Omega }}\left( k\rightarrow k_c\right) $
control the scaling properties as $t\rightarrow \infty $\cite{2,Kandel}.
Hence, (\ref{z}) again determines the dynamic exponent, merely $\lambda
/\omega $ should be replaced by $\lambda _{\max }/\omega _{\min }$\cite
{Kandel}, i.e. 
\begin{equation}
z=\frac{\ln \left( \lambda _{\max }/\omega _{\min }\right) }{\ln L} 
\eqnum{2.18}  \label{z*}
\end{equation}
where $\lambda _{\max }$ is the largest eigenvalue of the matrix ${\bf R}_{%
{\bf \Lambda }}\left( k\right) $, and $\omega _{\min }$ is the smallest
eigenvalue of ${\bf R}_{{\bf \Omega }}\left( k\right) .$

\section{Kinetic Gaussian model on three different fractal geometries}

\label{Sec.3}

\subsection{The Koch curve{\rm ,} the modified Gaussian model and the master
equation}

The fractals\cite{Mandelbrot} that we are going to study are constructed by
an iterative procedure in which each segment of the object is replaced by a
generator. Fig.1 shows two different configurations of the Koch curves\cite
{Mandelbrot} including nonbranching, and branching Koch curve. In the
iteration, each stage of the iteration is described by a length-rescaling
factor $L$, and the number of the segments in the lattice, $N^{\prime }$,
increases to $N$ by a relation $N/N^{\prime }=L^{D_f}$, which defines the
fractal dimensionality $D_f$. Obviously, these examples in Fig.1 have
different $D_f$, but their topological dimensionality $D_T$ is equal to $1$
which means that they are quasilinear fractals.. Added to this, another
parameter which is used to characterize the topological properties of the
fractal is $R$, the order of ramification. The maximum and minimum values of 
$R$ of a fractal obey the inequality, $R_{\max }\geq 2R_{\min }-2$\cite
{Gefen}$.$

The examples of the $D_T=1$ fractals shown in Fig.1 have finite $R$. The
nonbranching Koch curve (NBKC) which has $D_f=\ln 4/\ln 3$ and $R_{\min
}=R_{\max }=2$ shown in Fig.1(a) is a homogeneous and wiggling chain, while
the branching Koch curve (BKC) which has $D_f=\ln 5/\ln 3$ and $R_{\min }=2$
and $R_{\max }=3$ shown in Fig.1(b) is a inhomogeneous one.

We assume that the Gaussian spin system with a reduced Hamiltonian 
\begin{equation}
-\beta {\cal H}=k\sum_{\left\langle {i,j}\right\rangle }\sigma _i\sigma _j, 
\eqnum{3.1.1}  \label{3.1.1}
\end{equation}
located on these fractals, where $\beta =1/k_\beta T,k=J/k_\beta T$ and the
summation $\sum_{\left\langle {i,j}\right\rangle }{\text{ }}$is taken over
nearest neighbors. Unlike Ising spin system, the spin of the Gaussian model
can take any real value between ($-\infty ,+\infty $), and the Gaussian-type
distribution finding a given spin between $\sigma _i$ and $\sigma _i+d\sigma
_i$ 
\begin{equation}
f\left( \sigma _i\right) d\sigma _i\sim \exp \left( -\frac{b_{q_i}}2\sigma
_i^2\right) d\sigma _i  \eqnum{3.1.2}  \label{3.1.2}
\end{equation}
is assumed to prevent all spins from tending to infinity, where $q_i$ is the
coordination number of the site $i$, and $b_{q_i}$ is a distribution
constant independent of temperature. Considering the inhomogeneity of the
branching Koch curve, we have assumed that the Gaussian type distribution
constants depend on coordination numbers and satisfy a certain relation 
\begin{equation}
b_{q_i}/b_{q_j}=q_i/q_j.  \eqnum{3.1.3}  \label{3.1.3}
\end{equation}
As far as we know, this modified Gaussian model appeared firstly in Ref.\cite
{Kong-Li} which studied the static critical behavior of inhomogeneous
fractal lattices.

In this case the spin transition probability can be expressed as

\begin{equation}
W_i(\sigma _i\rightarrow \hat{\sigma}_i)=\frac 1{Q_i}\exp \left[ k\hat{\sigma%
}_i\sum_w\sigma _{i+w}\right] ,  \eqnum{3.1.4}  \label{3.1.4}
\end{equation}
where the normalized factor $Q_i$ can be determined as 
\[
Q_i=\sum_{\hat{\sigma}_i}\exp \left[ k\hat{\sigma}_i\sum_w\sigma
_{i+w}\right] =\int \exp \left[ k\hat{\sigma}_i\sum_w\sigma _{i+w}\right]
f\left( \hat{\sigma}_i\right) d\hat{\sigma}_i=\exp \left[ -\frac{k^2}{%
2b_{q_i}}\left( \sum_w\sigma _{i+w}\right) ^2\right] , 
\]
and the another useful combination formula can also be obtained 
\begin{eqnarray}
\sum_{\hat{\sigma}_i}\left( \sigma _i-\hat{\sigma}_i\right) W_i(\sigma _i
&\rightarrow &\hat{\sigma}_i)=\int_{-\infty }^\infty \left( \sigma _i-\hat{%
\sigma}_i\right) W_i(\sigma _i\rightarrow \hat{\sigma}_i)f(\hat{\sigma}_i)d%
\hat{\sigma}_i  \nonumber \\
&=&\sigma _i-\frac k{b_{q_i}}\sum_w\sigma _{i+w}.  \eqnum{3.1.5}
\label{3.1.5}
\end{eqnarray}
So, for magnetic-like perturbation, the master equation suitable for
modified Gaussian model on homogeneous and inhomogeneous fractal lattices
can be written as 
\begin{equation}
\frac d{dt}\sum_ih_{q_i}\left( t\right) \sigma _iP_{{\normalsize eq}}\left(
k,\left\{ \sigma \right\} \right) =-\sum_ih_{q_i}\left( t\right) \left(
\sigma _i-\frac k{b_{q_i}}\sum_w\sigma _{i+w}\right) P_{{\normalsize eq}%
}\left( k,\left\{ \sigma \right\} \right) .  \eqnum{3.1.6}  \label{3.1.6}
\end{equation}

\subsection{Nonbranching Koch curve}

First let us focus on the homogeneous nonbranching Koch curve (NBKC) in
which the Gaussian spins are placed on the all of sites. Because $%
h_{q_j}\left( t\right) =h\left( t\right) ,\ b_{q_j}=b,$ the master equation (%
\ref{3.1.6}) takes the following form 
\begin{eqnarray}
&&\left( \frac d{dt}\right) h\left( t\right) \sum_\alpha \left( \frac 12%
\sigma _1^\alpha +\sigma _2^\alpha +\sigma _3^\alpha +\sigma _4^\alpha +%
\frac 12\sigma _5^\alpha \right) P_{{\normalsize eq}}\left( k,\left\{ \sigma
\right\} \right)  \nonumber \\
&=&-h\left( t\right) \left( 1-\frac{2k}b\right) \sum_\alpha \left( \frac 12%
\sigma _1^\alpha +\sigma _2^\alpha +\sigma _3^\alpha +\sigma _4^\alpha +%
\frac 12\sigma _5^\alpha \right) P_{{\normalsize eq}}\left( k,\left\{ \sigma
\right\} \right) ,  \eqnum{3.2.1}  \label{3.2.1}
\end{eqnarray}
where the mark $\alpha $ denotes generator of NBKC which is shown in Fig.
2(a), the sum $\sum_\alpha $ goes over all generators, and $P_{eq}\left(
k,\left\{ \sigma \right\} \right) $ is the equilibrium distribution function
which can be written as 
\begin{eqnarray}
P_{{\normalsize eq}}\left( k,\left\{ \sigma \right\} \right) &=&\frac 1Z\exp
\left[ k\sum_{\left\langle i,j\right\rangle }\sigma _i\sigma _j-\frac b2%
\sum_i\sigma _i^2\right]  \nonumber \\
&=&\frac 1Z\prod_\alpha \exp \left\{ k\left( \sigma _1^\alpha \sigma
_2^\alpha +\sigma _2^\alpha \sigma _3^\alpha +\sigma _3^\alpha \sigma
_4^\alpha +\sigma _4^\alpha \sigma _5^\alpha \right) \right.  \nonumber \\
&&\left. -\frac b2\left[ \frac 12\left( \sigma _1^\alpha \right) ^2+\left(
\sigma _2^\alpha \right) ^2+\left( \sigma _3^\alpha \right) ^2+\left( \sigma
_4^\alpha \right) ^2+\frac 12\left( \sigma _5^\alpha \right) ^2\right]
\right\} .  \eqnum{3.2.2}  \label{3.2.2}
\end{eqnarray}
In Eqs. (\ref{3.2.1}) and (\ref{3.2.2}), the coefficient $1/2$ comes from
the fact that two neighboring generators share the same site $1$ and $5.$

The space-rescaling procedure (see Fig.3(a)) is performed through the
decimation renormalization transformation 
\begin{equation}
T^\alpha \left( \mu ,\sigma \right) =\delta \left( \mu _1^\alpha -\sigma
_1^\alpha \right) \delta \left( \mu _2^\alpha -\sigma _5^\alpha \right) . 
\eqnum{3.2.3}  \label{3.2.3}
\end{equation}
In which the spins $\sigma _1^\alpha $, $\sigma _5^\alpha $ and the
interaction $k$ in the $\alpha $-th generator are replaced by rescaling spin 
$\mu _1^{\prime \alpha }$, $\mu _2^\alpha $ and new interaction $k^{\prime }$%
, respectively, while the other spins $\sigma _2^\alpha ,$ $\sigma _3^\alpha 
$ and $\sigma _4^\alpha $ are integrated from $-\infty $ to $+\infty .$
Under this process, the form of the distribution function $P_{eq}$ is
invariant. The details of RG calculation is emplaced in appendix (\ref{app-a}%
). Here, we give a renormalized master equation 
\begin{eqnarray}
&&\left( \frac d{dt}\right) \frac 1{\xi \left( k\right) }\frac{b\left(
b+2k\right) }{b^2-2k^2}h\left( t\right) \sum_\alpha \left( \frac 12\mu
_1^{\prime \alpha }+\frac 12\mu _2^{\prime \alpha }\right) P_{{\normalsize eq%
}}^{\prime }\left( k^{\prime },\left\{ \mu ^{\prime }\right\} \right) 
\nonumber \\
&=&-\frac 1{\xi \left( k\right) }\frac{b^2-4k^2}{b^2-2k^2}h\left( t\right)
\sum_\alpha \left( \frac 12\mu _1^{\prime \alpha }+\frac 12\mu _2^{\prime
\alpha }\right) P_{{\normalsize eq}}^{\prime }\left( k^{\prime },\left\{ \mu
^{\prime }\right\} \right) ,  \eqnum{3.2.4}  \label{3.2.4}
\end{eqnarray}
where 
\begin{equation}
\mu ^{\prime }=\xi \left( k\right) \mu =\left( \frac{b^4-4k^2b^2+2k^4}{%
b^2\left( b^2-2k^2\right) }\right) ^{1/2}\mu ,  \eqnum{3.2.5}  \label{3.2.5}
\end{equation}
\begin{equation}
k^{\prime }=\frac{k^3b}{b^4-4k^2b^2+2k^4}k.  \eqnum{3.2.6}  \label{3.2.6}
\end{equation}
Obviously, if the summation for $\alpha $ is arranged in next stage of
iteration, the Eq. (\ref{3.2.4}) can be rewritten as 
\begin{eqnarray}
&&\left( \frac d{dt}\right) \frac 1{\xi \left( k\right) }\frac{b\left(
b+2k\right) }{b^2-2k^2}h\left( t\right) \sum_\beta \left( \frac 12\mu
_1^{\prime \beta }+\mu _2^{\prime \beta }+\mu _3^{\prime \beta }+\mu
_4^{\prime \beta }+\frac 12\mu _5^{\prime \beta }\right) P_{{\normalsize eq}%
}^{\prime }\left( k^{\prime },\left\{ \mu ^{\prime }\right\} \right) 
\nonumber \\
&=&-\frac 1{\xi \left( k\right) }\frac{b^2-4k^2}{b^2-2k^2}h\left( t\right)
\sum_\beta \left( \frac 12\mu _1^{\prime \beta }+\mu _2^{\prime \beta }+\mu
_3^{\prime \beta }+\mu _4^{\prime \beta }+\frac 12\mu _5^{\prime \beta
}\right) P_{{\normalsize eq}}^{\prime }\left( k^{\prime },\left\{ \mu
^{\prime }\right\} \right) .  \eqnum{3.2.7}  \label{3.2.7}
\end{eqnarray}
Futhermore, if we let 
\begin{equation}
\lambda =\frac 1{\xi \left( k\right) }\frac{b\left( b+2k\right) }{b^2-2k^2}%
,\ \ \omega =\frac 1{\xi \left( k\right) }\frac{b^2-4k^2}{b^2-2k^2}\cdot 
\frac 1{1-\frac{2k^{\prime }}b},  \eqnum{3.2.8}  \label{3.2.8}
\end{equation}
then by time rescaling 
\begin{equation}
t^{\prime }=\frac t{\lambda /\omega }=\frac{b^4-4k^2b^2+2k^4}{\left(
b+2k\right) b^3}t=L^{-z}t,\ \ (L=3)  \eqnum{3.2.9}  \label{3.2.9}
\end{equation}
and dynamic parameter transformation 
\begin{equation}
h\left( t\right) \rightarrow h^{\prime }\left( t^{\prime }\right) =\lambda
h\left( t\right) ,  \eqnum{3.2.10}  \label{3.2.10}
\end{equation}
the invariant form of the master equation (\ref{3.2.2}) can be restored 
\begin{eqnarray}
&&\left( \frac d{dt^{\prime }}h^{\prime }\left( t^{\prime }\right) \right)
\sum_\beta \left( \frac 12\mu _1^{\prime \beta }+\mu _2^{\prime \beta }+\mu
_3^{\prime \beta }+\mu _4^{\prime \beta }+\frac 12\mu _5^{\prime \beta
}\right) P_{{\normalsize eq}}^{\prime }\left( k^{\prime },\left\{ \mu
^{\prime }\right\} \right)  \nonumber \\
&=&-h^{\prime }\left( t^{\prime }\right) \left( 1-\frac{2k^{\prime }}b%
\right) \sum_\beta \left( \frac 12\mu _1^{\prime \beta }+\mu _2^{\prime
\beta }+\mu _3^{\prime \beta }+\mu _4^{\prime \beta }+\frac 12\mu _5^{\prime
\beta }\right) P_{{\normalsize eq}}^{\prime }\left( k^{\prime },\left\{ \mu
^{\prime }\right\} \right) .  \eqnum{3.2.11}  \label{3.2.11}
\end{eqnarray}

Let system is in its critical point $k_c=b/2$ which determined by the
recursion relationship (\ref{3.2.6}), then we can obtain the dynamical
critical exponent $z$ by use of Eqs. (\ref{3.2.9}) as

\begin{equation}
z=\left[ \frac 1{\ln L}\ln \frac{\left( b+2k\right) b^3}{b^4-4k^2b^2+2k^4}%
\right] _{k_c=b/2,L=3}=2\frac{\ln 4}{\ln 3}=2D_f.  \eqnum{3.2.12}
\label{3.2.12}
\end{equation}
However, because 
\begin{eqnarray}
\frac 1\nu &=&\left. \frac 1{\ln L}\ln \left( \frac{dk^{\prime }}{dk}\right)
\right| _{k_c=b/2,L=3}=\left. \frac 1{\ln L}\ln \left( \frac{-4k^3b^3\left(
-b^2+2k^2\right) }{\left( b^4-4k^2b^2+2k^4\right) ^2}\right) \right|
_{k_c=b/2,L=3}  \nonumber \\
&=&\frac{\ln 16}{\ln 3}=2\frac{\ln 4}{\ln 3}=2D_f.  \eqnum{3.2.13}
\label{3.2.13}
\end{eqnarray}
then 
\begin{equation}
z=\frac 1\nu =2D_f=2.\,5237.  \eqnum{3.2.14}  \label{3.2.14}
\end{equation}

\subsection{Branching Koch curve}

Now, we turn to focus on the branching Koch curve (BKC) which is one of the
inhomogeneous fractals. In this case, due to the coordination number depends
on the place of site, we must to assume that the Gaussian type distribution
constants depend on coordination numbers and satisfy a certain relation (\ref
{3.1.3}), otherwise, the problem can not be solved by applying the
decimation RG method directly\cite{Kong-Li}.

In the following we deal with the magnetic-like perturbation master equation
that suits for modified Gaussian model on inhomogeneous fractal lattices (%
\ref{3.1.6}). We should notice that, for the branching Koch curve (BKC)
there are two kinds of typical generators (such as $\alpha $-th and $\beta $%
-th in Fig.2(b)): (1) $q_1=q_2=q_4=3$, $q_3=q_5=2$; (2) $q_1=q_2=q_4=q_5=3$, 
$q_3=2$ $.$ For case (1) or case (2), the decimation renormalization group
procedure is shown in figure 3(b), in which some spins such as $\sigma _2,$ $%
\sigma _3$ and $\sigma _4$ are integrated, the remainders are rescaled as $%
\mu _1^{\prime }$ and $\mu _2^{\prime }$, and, at the same time, the
interaction $k$ is replaced by $k^{\prime }.$ Under this process, the form
of the distribution function is invariant, and the RG transformation of $%
\alpha $-th generator is equivalent to $\beta $-th. It can be realized via
the calculation of the appendix (\ref{app-b1}).

Our purposes is the renormalization of the master equation. In fact, we only
need discuss a typical generator. Without loss of generality, we take the $%
\alpha $-th generator, for instance. The lift and right sides of Eq.(\ref
{3.1.6}) can be written as, respectively 
\begin{equation}
{\bf h}\left( t\right) {\bf \cdot \Lambda \left( \sigma \right) }P_{%
{\normalsize eq}}\left( k,\left\{ \sigma \right\} \right) =\left[ h_3\left(
t\right) \left( \frac 13\sigma _1^\alpha +\sigma _2^\alpha +\sigma _4^\alpha
\right) +h_2\left( t\right) \left( \sigma _3^\alpha +\frac 12\sigma
_5^\alpha \right) \right] P_{{\normalsize eq}}\left( k,\left\{ \sigma
\right\} \right) ,  \eqnum{3.3.1}  \label{3.3.1}
\end{equation}
\begin{eqnarray}
&&{\bf h}\left( t\right) {\bf \cdot \Omega }\left( k,\sigma \right) P_{%
{\normalsize eq}}\left( k,\left\{ \sigma \right\} \right)  \nonumber \\
&=&\left\{ h_3\left[ \frac 13\sigma _1^\alpha -\frac k{b_3}\sigma _2^\alpha
\right] +h_3\left[ \sigma _2^\alpha -\frac k{b_3}\left( \sigma _1^\alpha
+\sigma _3^\alpha +\sigma _4^\alpha \right) \right] +h_2\left[ \sigma
_3^\alpha -\frac k{b_2}\left( \sigma _2^\alpha +\sigma _4^\alpha \right)
\right] \right.  \nonumber \\
&&\left. +h_3\left[ \sigma _4^\alpha -\frac k{b_3}\left( \sigma _2^\alpha
+\sigma _3^\alpha +\sigma _5^\alpha \right) \right] +h_2\left[ \frac 12%
\sigma _5^\alpha -\frac k{b_2}\sigma _4^\alpha \right] \right\} P_{%
{\normalsize eq}}\left( k,\left\{ \sigma \right\} \right) ,  \eqnum{3.3.2}
\label{3.3.2}
\end{eqnarray}
where, the coefficient $1/3$ (or $1/2$) in the terms $\sigma _1^\alpha $ (or 
$\sigma _5^\alpha $ ) comes from the fact that three (or two) neighboring
generators share the same site $1$ (or $5$ ).

Multiplying Eqs. (\ref{3.3.1}) and (\ref{3.3.2}) by the transformation
operator 
\begin{equation}
T\left( \mu ,\sigma \right) =\prod_\alpha \delta \left( \mu _1^\alpha
-\sigma _1^\alpha \right) \delta \left( \mu _2^\alpha -\sigma _5^\alpha
\right) ,  \eqnum{3.3.3}  \label{3.3.3}
\end{equation}
and integrate over the $\left\{ \sigma \right\} $, one can obtain (see
appendix (\ref{app-b2})) 
\begin{eqnarray}
&&R\left\{ {\bf h}\left( t\right) {\bf \cdot \Lambda }\left( \sigma \right)
P_{{\normalsize eq}}\left( k,\left\{ \sigma \right\} \right) \right\} 
\nonumber \\
&=&\frac 1{\sqrt{\xi }}\left( 
\begin{array}{ll}
h_3^{\prime }\left( t\right) & h_2^{\prime }\left( t\right)
\end{array}
\right) \left( 
\begin{array}{l}
\frac 13\mu _1^{\prime \alpha } \\ 
\frac 12\mu _2^{\prime \alpha }
\end{array}
\right) P_{{\normalsize eq}}^{\prime }\left( \left\{ k^{\prime },\mu
^{\prime }\right\} \right)  \nonumber \\
&=&{\bf h}^{\prime }\left( t,k^{\prime }\right) {\bf \cdot \Lambda }^{\prime
}\left( \mu ^{\prime },k^{\prime }\right) P_{{\normalsize eq}}^{\prime
}\left( k^{\prime },\left\{ \mu ^{\prime }\right\} \right) ,  \eqnum{3.3.4}
\label{3.3.4}
\end{eqnarray}
\begin{eqnarray}
&&R\left\{ {\bf h}\left( t\right) {\bf \cdot \Omega }\left( k,\sigma \right)
P_{{\normalsize eq}}\left( k,\left\{ \sigma \right\} \right) \right\} 
\nonumber \\
&=&\frac 1{\sqrt{\xi }}\left( 
\begin{array}{ll}
h_3^{\prime }\left( t\right) & h_2^{\prime }\left( t\right)
\end{array}
\right) \frac{\left( R_\Omega \right) }{\left( R_\Lambda \right) }\left( 
\begin{array}{l}
\frac 13\mu _1^{\prime \alpha }-\frac{k^{\prime }}{b_3}\mu _2^{\prime \alpha
} \\ 
\frac 12\mu _2^{\prime \alpha }-\frac{k^{\prime }}{b_2}\mu _1^{\prime \alpha
}
\end{array}
\right) P_{{\normalsize eq}}^{\prime }\left( \left\{ k^{\prime },\mu
^{\prime }\right\} \right)  \nonumber \\
&=&{\bf h}^{\prime }\left( t,k^{\prime }\right) {\bf \cdot }\left[ {\bf %
R_\Lambda ^{-1}\left( k\right) \cdot R_\Omega \left( k\right) }\right] {\bf %
\cdot \Omega }^{\prime }\left( \mu ^{\prime },k^{\prime }\right) P_{%
{\normalsize eq}}^{\prime }\left( k^{\prime },\left\{ \mu ^{\prime }\right\}
\right) ,  \eqnum{3.3.5}  \label{3.3.5}
\end{eqnarray}
in which 
\[
\left( 
\begin{array}{ll}
h_3^{\prime }\left( t\right) & h_2^{\prime }\left( t\right)
\end{array}
\right) =\left( 
\begin{array}{ll}
h_3\left( t\right) & h_2\left( t\right)
\end{array}
\right) \left( R_\Lambda \right) , 
\]
\begin{eqnarray}
\left( R_\Lambda \right) _{k\rightarrow k_c} &=&\left( 
\begin{array}{cc}
\frac{b_3b_2+2kb_2-2k^2}{b_3b_2-kb_2-2k^2} & \frac{2kb_2}{b_3b_2-kb_2-2k^2}
\\ 
\frac{3k^2}{b_3b_2-kb_2-2k^2} & \frac{b_2\left( b_3-k\right) }{%
b_3b_2-kb_2-2k^2}
\end{array}
\right) _{k\rightarrow k_c}  \nonumber \\
&=&\left( 
\begin{array}{cc}
4 & 2 \\ 
\frac 32 & 2
\end{array}
\right) ,{\normalsize eigenvalues}:5,1,  \eqnum{3.3.6}  \label{3.3.6}
\end{eqnarray}
\begin{eqnarray}
\left( R_\Omega \right) _{k\rightarrow k_c} &=&\left( 
\begin{array}{cc}
\frac{9b_2^4-28k^2b_2^2-8k^3b_2+8k^4}{\left( 9b_2^3-16k^2b_2-8k^3\right) b_2}
& 0 \\ 
0 & \frac{9b_2^4-28k^2b_2^2-8k^3b_2+8k^4}{\left( 9b_2^3-16k^2b_2-8k^3\right)
b_2}
\end{array}
\right) _{k\rightarrow k_c}  \nonumber \\
&=&\left( 
\begin{array}{cc}
\frac 38 & 0 \\ 
0 & \frac 38
\end{array}
\right) ,{\normalsize eigenvalues}:\frac 38,\frac 38,  \eqnum{3.3.7}
\label{3.3.7}
\end{eqnarray}
where $k_c=b_2/2=b_3/3$ is determined by the fixed-point equation $%
k^{*}=k^{\prime }=k$ ( see appendix (\ref{app-b1})).

In this case we have to look for the invariant form of the master equation 
\begin{equation}
\frac d{dt}{\bf h}\left( t\right) {\bf \cdot \Lambda }\left( \sigma \right)
P_{{\normalsize eq}}\left( k,\left\{ \sigma \right\} \right) =-{\bf h}\left(
t\right) {\bf \cdot \Omega }\left( k,\sigma \right) P_{{\normalsize eq}%
}\left( k,\left\{ \sigma \right\} \right)  \eqnum{3.3.8}  \label{4.3.8}
\end{equation}
at the limit of the order $n\rightarrow \infty $ of the RG transformation.
Because our starting point is very close to the fixed point $k_c$ of the
static RG transformation, the eigenvalues of the transformation matrices $%
{\bf R}_{{\bf \Lambda }}\left( k\rightarrow k_c\right) $ and ${\bf R}_{{\bf %
\Omega }}\left( k\rightarrow k_c\right) $ control the scaling properties of
the largest relaxation time$.$ Hence, according to the foregoing discussion,
the dynamical critical exponent $z$ is obtained 
\begin{equation}
z=\frac{\ln \left( \lambda _{\max }/\omega _{\min }\right) }{\ln L}=\frac{%
\ln 40/3}{\ln 3}=2.\,3578.  \eqnum{3.3.9}  \label{3.3.9}
\end{equation}
Because of 
\[
\left( \frac{dk^{\prime }}{dk}\right) _{k=k_c}=\frac d{dk}\left( \frac{%
4b_2k^3\left( k+b_2\right) }{8k^4-8k^3b_2-28b_2^2k^2+9b_2^4}\right) _{k=k_c}=%
\frac{40}3, 
\]
\begin{equation}
\frac 1\nu =\frac 1{\ln L}\ln \left( \frac{dk^{\prime }}{dk}\right) _{k=k_c}=%
\frac{\ln 40/3}{\ln 3},  \eqnum{3.3.10}  \label{3.3.10}
\end{equation}
then 
\begin{equation}
z=\frac 1\nu =2.\,3578.  \eqnum{3.3.11}  \label{3.3.11}
\end{equation}

\subsection{Multibranching Koch Curve}

Based on the branching Koch curve, we now construct another generalized one,
the multibranching Koch curve (MBKC), and investigate its critical dynamical
behavior of the kinetic modified Gaussian model on this lattice. The
constructional process is shown in Fig.4. Obviously, it also is a
inhomogeneous example of the $D_T=1$ fractals which has $D_f=\ln (2m+3)/\ln
3,$ $R_{\min }=2$ and $R_{\max }=m+2,$ $m=1,...\infty .$ The effect of the
order of ramification, $R,$ on the critical slowing down could be seen by
this example.

Similarly, there are two kinds of typical generators in the multibranching
Koch curve: (1) $q_1=q_3=q_4=1/(m+2),$ $q_i=2,(i=5,\cdots ,m+4),$ $q_2=2$ ( $%
\alpha -$th generator); (2) $q_1=q_3=q_4=q_2=1/(m+2),$ $q_i=2,(i=5,\cdots
,m+4)$ ( $\beta -$th generator). In fact, the decimation renormalizing
procedures of these two cases result in the same consequence (see Fig.5). It
can be realized via the calculation of the appendix (\ref{app-c1}).

Our purposes is the renormalization of the master equation 
\begin{equation}
\frac d{dt}\left( \sum_ih_{q_i}\left( t\right) \sigma _i\right) P_{%
{\normalsize eq}}\left( k,\left\{ \sigma \right\} \right)
=-\sum_ih_{q_i}\left( t\right) \left( \sigma _i-\frac k{b_{q_i}}\sum_w\sigma
_{i+w}\right) P_{{\normalsize eq}}\left( k,\left\{ \sigma \right\} \right) ,
\eqnum{3.4.1}  \label{3.4.1}
\end{equation}
or 
\begin{equation}
\frac d{dt}{\bf h}\left( t\right) {\bf \cdot \Lambda }\left( \sigma \right)
P_{{\normalsize eq}}\left( k,\left\{ \sigma \right\} \right) =-{\bf h}\left(
t\right) {\bf \cdot \Omega }\left( k,\sigma \right) P_{{\normalsize eq}%
}\left( k,\left\{ \sigma \right\} \right) .  \eqnum{3.4.2}  \label{3.4.2}
\end{equation}
In fact, we only need discuss a typical generator. Without loss of
generality, we take case (1) for instance. The lift and right sides of Eq.(%
\ref{3.4.1}) can be written as, respectively 
\begin{equation}
{\bf h}\left( t\right) {\bf \cdot \Lambda }\left( \sigma \right) P_{%
{\normalsize eq}}\left( k,\left\{ \sigma \right\} \right) =\left[
h_{m+2}\left( \frac 1{m+2}\sigma _1^\alpha +\sigma _3^\alpha +\sigma
_4^\alpha \right) +h_2\left( \sum_{i=5}^{m+4}\sigma _i^\alpha +\frac 12%
\sigma _2^\alpha \right) \right] P_{{\normalsize eq}}\left( k,\left\{ \sigma
\right\} \right)  \eqnum{3.4.3}  \label{3.4.3}
\end{equation}
and 
\begin{eqnarray}
&&{\bf h}\left( t\right) {\bf \cdot \Omega }\left( k,\sigma \right) P_{%
{\normalsize eq}}\left( k,\left\{ \sigma \right\} \right)  \nonumber \\
&=&\left\{ h_{m+2}\left[ \frac 1{m+2}\sigma _1^\alpha -\frac k{b_{m+2}}%
\sigma _3^\alpha \right] +h_{m+2}\left[ \sigma _3^\alpha -\frac k{b_{m+2}}%
\left( \sigma _1^\alpha +\sum_{i=5}^{m+4}\sigma _i^\alpha +\sigma _4^\alpha
\right) \right] \right.  \nonumber \\
&&+h_{m+2}\left[ \sigma _4^\alpha -\frac k{b_{m+2}}\left( \sigma _3^\alpha
+\sum_{i=5}^{m+4}\sigma _i^\alpha +\sigma _2^\alpha \right) \right] 
\nonumber \\
&&\left. +\sum_{i=5}^{m+4}h_2\left[ \sigma _i^\alpha -\frac k{b_2}\left(
\sigma _3^\alpha +\sigma _4^\alpha \right) \right] +h_2\left[ \frac 12\sigma
_2^\alpha -\frac k{b_2}\sigma _4^\alpha \right] \right\} P_{{\normalsize eq}%
}\left( k,\left\{ \sigma \right\} \right) .  \eqnum{3.4.4}  \label{3.4.4}
\end{eqnarray}
By virtue of (\ref{C.2.1}-\ref{C.2.5}), the results of the decimation RG
transformation of (\ref{3.4.3}) and (\ref{3.4.4}) are, respectively 
\begin{eqnarray}
&&R\left\{ {\bf h}\left( t\right) {\bf \cdot \Lambda }\left( \sigma \right)
P_{{\normalsize eq}}\left( k,\left\{ \sigma \right\} \right) \right\} 
\nonumber \\
&=&\frac 1{\sqrt{\xi }}\left( 
\begin{array}{ll}
h_{m+2}^{\prime }\left( t,k^{\prime }\right) & h_2^{\prime }\left(
t,k^{\prime }\right)
\end{array}
\right) \left( 
\begin{array}{l}
\frac 1{m+2}\mu _1^{\prime \alpha } \\ 
\frac 12\mu _2^{\prime \alpha }
\end{array}
\right) P_{{\normalsize eq}}^{\prime }\left( k^{\prime },\left\{ \mu
^{\prime }\right\} \right) ,  \nonumber \\
&=&{\bf h}^{\prime }\left( t,k^{\prime }\right) {\bf \cdot \Lambda }^{\prime
}\left( \mu ^{\prime },k^{\prime }\right) P_{{\normalsize eq}}^{\prime
}\left( k^{\prime },\left\{ \mu ^{\prime }\right\} \right)  \eqnum{3.4.5}
\label{3.4.5}
\end{eqnarray}
where 
\[
\left( 
\begin{array}{ll}
h_{m+2}^{\prime } & h_2^{\prime }
\end{array}
\right) =\left( 
\begin{array}{ll}
h_{m+2} & h_2
\end{array}
\right) \left( R_\Lambda \right) 
\]
\[
\left( R_\Lambda \right) =\left( 
\begin{array}{ll}
\frac{b_2b_{m+2}-b_2k-2mk^2+kb_2\left( m+2\right) }{\left(
b_2b_{m+2}-b_2k-2mk^2\right) } & \frac{2b_2k}{b_2b_{m+2}-b_2k-2mk^2} \\ 
\frac{\left( m+2\right) mk^2}{b_2b_{m+2}-b_2k-2mk^2} & \frac{b_2b_{m+2}-b_2k%
}{b_2b_{m+2}-b_2k-2mk^2}
\end{array}
\right) 
\]
and 
\begin{eqnarray}
&&R\left\{ {\bf h}\left( t\right) {\bf \cdot \Omega }\left( k,\sigma \right)
P_{{\normalsize eq}}\left( k,\left\{ \sigma \right\} \right) \right\} 
\nonumber \\
&=&\frac 1{\sqrt{\xi }}\left( 
\begin{array}{ll}
h_{m+1}^{\prime }\left( t,k^{\prime }\right) & h_2^{\prime }\left(
t,k^{\prime }\right)
\end{array}
\right) \frac{\left( R_\Omega \right) }{\left( R_\Lambda \right) }\left( 
\begin{array}{l}
\frac 1{m+2}\mu _1^{\prime \alpha }-\frac{k^{\prime }}{b_{m+2}}\mu
_2^{\prime \alpha } \\ 
-\frac{k^{\prime }}{b_2}\mu _1^{\prime \alpha }+\frac 12\mu _2^{\prime
\alpha }
\end{array}
\right) P_{{\normalsize eq}}^{\prime }\left( k^{\prime },\left\{ \mu
^{\prime }\right\} \right) ,  \nonumber \\
&=&{\bf h}^{\prime }\left( t,k^{\prime }\right) {\bf \cdot }\left[ {\bf %
R_\Lambda ^{-1}\left( k\right) \cdot R_\Omega \left( k\right) }\right] {\bf %
\cdot \Omega }^{\prime }\left( \mu ^{\prime },k^{\prime }\right) P_{%
{\normalsize eq}}^{\prime }\left( k^{\prime },\left\{ \mu ^{\prime }\right\}
\right)  \eqnum{3.4.6}  \label{3.4.6}
\end{eqnarray}
where 
\[
\left( R_\Omega \right) _{k=k_c}=\left( \left( \tilde{R}_\Omega \right)
\left( 
\begin{array}{ll}
1 & -\frac{2k^{\prime }}{b_{m+2}} \\ 
-\frac{\left( m+2\right) k^{\prime }}{b_2} & 1
\end{array}
\right) ^{-1}\right) ,\quad \left( \tilde{R}_\Omega \right) =\left( 
\begin{array}{ll}
a_{11} & a_{12} \\ 
a_{21} & a_{22}
\end{array}
\right) , 
\]
\[
a_{11}=\frac{-2b_{m+2}mk^3-b_{m+2}b_2k^2-2b_{m+2}^2mk^2+b_2b_{m+2}^3-b_2k^2%
\left( m+2\right) b_{m+2}+k^4\left( m+2\right) m}{\left(
b_2b_{m+2}-b_2k-2mk^2\right) b_{m+2}\left( k+b_{m+2}\right) }, 
\]
\[
a_{12}=\frac{2k^3\left( mk+b_2\right) }{b_{m+2}\left( k+b_{m+2}\right)
\left( 2mk^2+b_2k-b_2b_{m+2}\right) }, 
\]
\[
a_{21}=\frac{\left( m+2\right) k^3\left( mk+b_2\right) }{b_2\left(
k+b_{m+2}\right) \left( 2mk^2+b_2k-b_2b_{m+2}\right) }, 
\]
\[
a_{22}=\frac{%
2b_2mk^3+b_2^2k^2+2b_2b_{m+2}mk^2-b_2^2b_{m+2}^2+2b_{m+2}b_2k^2-2k^4m}{%
b_2\left( k+b_{m+2}\right) \left( 2mk^2+b_2k-b_2b_{m+2}\right) }, 
\]
\begin{equation}
\left( R_\Omega \right) _{k\rightarrow k_c}=\left( 
\begin{array}{cc}
\frac{m+2}{2\left( m+3\right) } & 0 \\ 
0 & \frac{m+2}{2\left( m+3\right) }
\end{array}
\right) ,{\normalsize eigenvalues}:\frac{m+2}{2\left( m+3\right) },\frac{m+2%
}{2\left( m+3\right) }  \eqnum{3.4.7}  \label{3.4.7}
\end{equation}
\begin{equation}
\left( R_\Lambda \right) _{k\rightarrow k_c}=\left( 
\begin{array}{cc}
m+3 & 2 \\ 
\frac 12m\left( m+2\right) & m+1
\end{array}
\right) ,{\normalsize eigenvalues}:2m+3,1  \eqnum{3.4.8}  \label{3.4.8}
\end{equation}
where $k_c$ is determined by the fixed-point equation $k^{*}=k^{\prime }=k$
(see (\ref{C.1.12})) as 
\begin{equation}
k_c=\frac{b_2}2=\frac{b_{m+2}}{m+2}.  \eqnum{3.4.9}  \label{3.4.9}
\end{equation}

From Eqs. (\ref{3.4.5}) and (\ref{3.4.6}) we can see that, in order to look
for the invariant form of the master equation 
\begin{equation}
\frac d{dt}{\bf h}\left( t\right) {\bf \cdot \Lambda }\left( \sigma \right)
P_{{\normalsize eq}}\left( k,\left\{ \sigma \right\} \right) =-{\bf h}\left(
t\right) {\bf \cdot \Omega }\left( k,\sigma \right) P_{{\normalsize eq}%
}\left( k,\left\{ \sigma \right\} \right)  \eqnum{3.4.10}  \label{3.4.10}
\end{equation}
we need to do the renormalization endlessly up to the limit of the order $%
n\rightarrow \infty $ of the RG transformation. But, because the system is
very close to the fixed point $k_c$ of the static RG transformation, the
scaling properties of the largest relaxation time are under the control of
the largest eigenvalue $\lambda _{\max }=2m+3$ of the matrix ${\bf R}_{{\bf %
\Lambda }}\left( k\rightarrow k_c\right) $ and the smallest eigenvalue $%
\omega _{\min }=\frac{m+2}{2\left( m+3\right) }$ of ${\bf R}_{{\bf \Omega }%
}\left( k\rightarrow k_c\right) .$ Hence, the invariant form of the master
equation can be obtained through preforming the time rescaling 
\begin{equation}
t\rightarrow t^{\prime }=L^{-z}t=\frac t{\lambda _{\max }/\omega _{\min }}, 
\eqnum{3.4.11}  \label{3.4.11}
\end{equation}
and from here, the dynamical critical exponent $z$ can be got

\begin{equation}
z=\frac{\ln \left( \lambda _{\max }/\omega _{\min }\right) }{\ln L}=\frac 1{%
\ln 3}\ln \frac{2\left( 2m+3\right) \left( m+3\right) }{m+2}.  \eqnum{3.4.12}
\label{3.4.12}
\end{equation}
Because of 
\begin{eqnarray*}
&&\left( \frac{dk^{\prime }}{dk}\right) _{k\rightarrow k_c} \\
&=&\frac d{dk}\left( \frac{4b_2k^3\left( b_2+mk\right) }{8k^4m-8b_2mk^3-4%
\left( m+2\right) b_2^2mk^2-4k^2b_2^2\left( m+2\right) -4k^2b_2^2+\left(
m+2\right) ^2b_2^4}\right) _{k\rightarrow b_2/2} \\
&=&2\frac{\left( m+3\right) \left( 2m+3\right) }{m+2}
\end{eqnarray*}
\begin{equation}
\frac 1\nu =\frac{\ln \left( \frac{dk^{\prime }}{dk}\right) _{k\rightarrow
k_c}}{\ln L}=\frac 1{\ln 3}\ln \frac{2\left( m+3\right) \left( 2m+3\right) }{%
m+2},  \eqnum{3.4.13}  \label{3.4.13}
\end{equation}
then 
\begin{equation}
z=\frac 1\nu =\frac 1{\ln 3}\ln \frac{2\left( m+3\right) \left( 2m+3\right) 
}{m+2}.  \eqnum{3.4.14}  \label{3.4.14}
\end{equation}

\section{Conclusions}

\label{Sec.4}

Based on the dynamical real-space renormalization proposed by Achiam and
Kosterlitz, we have suggested a generalizing formulation that suits
arbitrary spin systems. The new version replaces the single-spin flipping
Glauber dynamics with the single-spin transition dynamics. As an
application, we focused our minds on the kinetic Gaussian model ($\sigma
=-\infty ,...,\infty $, continuous spin model), and studied three different
fractal geometries with quasilinear lattices, including nonbranching,
branching and multibranching Koch curve. We calculated the dynamical
critical exponent $z$ for these lattices using an exact decimation
renormalization transformation in the assumption of the magnetic-like
perturbation, and found that it can be written universally as $z=1/\nu $,
where $\nu $ is the static length-correlation exponent.

In the first example, the nonbranching Koch curve, $z=1/\nu =2D_f$, $D_f=\ln
4/\ln 3$ is the fractal dimensionality of the NBKC. Being a quasilinear
chain, the geometrical effect of the wiggliness of the nonbranching Koch
curve is that the correlation length $\tilde{\zeta}$ of the one-dimensional
linear chain should be replaced by the real correlation length $\zeta ,$ $%
\zeta =\tilde{\zeta}^{1/D_f}$\cite{4}. However, for one-dimensional linear
chain with Gaussian spin each lattices, we have known that the critical
dynamical exponent $\bar{z}=2$\cite{Zhu-1}. So $\tau \sim \tilde{\zeta}^{%
\bar{z}}=\left( \zeta ^{D_f}\right) ^{\bar{z}}=\zeta ^z$ means $z=\bar{z}%
D_f=2D_f$. This result is coincided with our calculation by DRSRG technique.

In the branching Koch curve, the result $z=1/\nu $ is also valid, but that $%
\frac 1\nu =\frac 1{\ln L}\ln \left( \frac{dk^{\prime }}{dk}\right) _{k=k_c}=%
\frac{\ln 40/3}{\ln 3}$ is not simply related to the fractal dimensionality $%
D_f$.

In the multibranching Koch curve, the result $z=1/\nu $ is obtained once
again. We can see from $z=\frac 1{\ln 3}\ln \frac{2\left( m+3\right) \left(
2m+3\right) }{m+2}$ that, when $m=0$, $z=2$, the lattice is one-dimensional
chain and $z$ is equal to the rigorous result; when $m=1$, $z=\frac{\ln 40/3%
}{\ln 3}$, corresponding to the branching Koch curve; when $m\rightarrow
\infty $, $z\rightarrow \infty $ (see Fig.6). All of these means that the
critical slowing down of the Gaussian spin on Koch curve is dependent
mightily on the order of ramification $R.$

In fact, in our previous paper we have found that for a translational
symmetric lattices with Gaussian spin model, the critical dynamical exponent 
$z=1/\nu ,$ $\nu =1/2$ at the critical point $K_c=b/2d$ based on rigorous
calculation\cite{Zhu-1}. Yet, in this paper the result $z=1/\nu ,$ have been
proved once again by the dialational symmetric lattice systems. We guess
that $z=1/\nu ,$ could be a universal conclusion for kinetic Gaussian model.
Of course, we must realize that the result what we have obtained in this
paper is carried out in the assumption of the magnetic-like perturbation.
However, the perturbation itself (magnetic-like or energy-like) is only a
special assumption. For a general perturbation the master equation is not
always invariant under the RG transformation because the perturbations
probably have components along all the relevant operators. By this token,
whether the $z=1/\nu $ will be a universal conclusion for kinetic Gaussian
model waits for further investigation.

\section{Acknowledgments}

This work is supported by the National Basic Research Project ``Nonlinear
Science'', the National Natural Science Foundation of China under Grant No.
19775008 and Jiangxi Province Natural Science Foundation of China under
Grant No. 9912001. J. Y. Zhu thanks Dr. Z. Gao for his valuable discussions.

\appendix 

\section{RG calculation of NBKC}

\label{app-a}

To perform the decimation transformation, we have only to multiply both
sides of the master equation (\ref{3.2.1}) by the transformation operator 
\begin{equation}
T\left( \mu ,\sigma \right) =\prod_\beta \delta \left( \mu _1^\beta -\sigma
_1^\beta \right) \delta \left( \mu _2^\beta -\sigma _5^\beta \right) 
\eqnum{A.1}  \label{A.1}
\end{equation}
and integrate over the $\left\{ \sigma \right\} $, i.e. 
\begin{eqnarray}
&&\left( \frac d{dt}\right) h\left( t\right) \sum_\alpha R\left\{ \left( 
\frac 12\sigma _1^\alpha +\sigma _2^\alpha +\sigma _3^\alpha +\sigma
_4^\alpha +\frac 12\sigma _5^\alpha \right) P_{{\normalsize eq}}\left(
k,\left\{ \sigma \right\} \right) \right\}  \nonumber \\
&=&-h\left( t\right) \left( 1-\frac{2k}b\right) \sum_\alpha R\left\{ \left( 
\frac 12\sigma _1^\alpha +\sigma _2^\alpha +\sigma _3^\alpha +\sigma
_4^\alpha +\frac 12\sigma _5^\alpha \right) P_{{\normalsize eq}}\left(
k,\left\{ \sigma \right\} \right) \right\}  \eqnum{A.2}  \label{A.2}
\end{eqnarray}
where 
\begin{eqnarray}
&&R\left\{ \sigma _i^\alpha P_{{\normalsize eq}}\left( k,\left\{ \sigma
\right\} \right) \right\}  \nonumber \\
&=&\int_{-\infty }^\infty d\sigma _1d\sigma _2\ldots d\sigma _N\prod_\beta
\delta \left( \mu _1^\beta -\sigma _1^\beta \right) \delta \left( \mu
_2^\beta -\sigma _5^\beta \right) \sigma _i^\alpha P_{{\normalsize eq}%
}\left( k,\left\{ \sigma \right\} \right)  \nonumber \\
&=&=R\left\{ P_{{\normalsize eq}}\left( k,\left\{ \sigma \right\} \right)
\right\} \frac{W_{\sigma _i^\alpha }}W,  \eqnum{A.3}  \label{A.3}
\end{eqnarray}
\begin{eqnarray*}
W &=&\int_{-\infty }^\infty d\sigma _2^\alpha d\sigma _3^\alpha d\sigma
_4^\alpha \exp \left\{ k\left( \mu _1^\alpha \sigma _2^\alpha +\sigma
_2^\alpha \sigma _3^\alpha +\sigma _3^\alpha \sigma _4^\alpha +\sigma
_4^\alpha \mu _2^\alpha \right) \right. \\
&&\left. -\frac b2\left[ \left( \sigma _2^\alpha \right) ^2+\left( \sigma
_3^\alpha \right) ^2+\left( \sigma _4^\alpha \right) ^2\right] \right\} \\
&=&\sqrt{\frac{\left( 2\pi \right) ^3}{b\left( b^2-2k^2\right) }}\exp
\left\{ \frac{k^4}{b\left( b^2-2k^2\right) }\mu _1^\alpha \mu _2^\alpha +%
\frac 12\frac{k^2\left( b^2-k^2\right) }{b\left( b^2-2k^2\right) }\left[
\left( \mu _1^\alpha \right) ^2+\left( \mu _2^\alpha \right) ^2\right]
\right\} ,
\end{eqnarray*}
\begin{eqnarray*}
W_{\sigma _i^\alpha } &=&\int_{-\infty }^\infty d\sigma _2^\alpha d\sigma
_3^\alpha d\sigma _4^\alpha \cdot \sigma _i^\alpha \exp \left\{ k\left( \mu
_1^\alpha \sigma _2^\alpha +\sigma _2^\alpha \sigma _3^\alpha +\sigma
_3^\alpha \sigma _4^\alpha +\sigma _4^\alpha \mu _2^\alpha \right) \right. \\
&&\left. -\frac b2\left[ \left( \sigma _2^\alpha \right) ^2+\left( \sigma
_3^\alpha \right) ^2+\left( \sigma _4^\alpha \right) ^2\right] \right\} ,
\end{eqnarray*}
\[
W_{\sigma _1^\alpha }=\mu _1^\alpha W,\quad W_{\sigma _5^\alpha }=\mu
_2^\alpha W, 
\]
\[
W_{\sigma _2^\alpha }=\frac kb\left[ \mu _1^\alpha +\frac{k^2}{b^2-2k^2}%
\left( \mu _1^\alpha +\mu _2^\alpha \right) \right] W, 
\]
\[
W_{\sigma _3^\alpha }=\frac{k^2}{b^2-2k^2}\left( \mu _1^\alpha +\mu
_2^\alpha \right) W, 
\]
\[
W_{\sigma _4^\alpha }=\frac kb\left[ \mu _2^\alpha +\frac{k^2}{b^2-2k^2}%
\left( \mu _1^\alpha +\mu _2^\alpha \right) \right] W. 
\]

The remanent integration $R\left\{ P_{eq}\left( k,\left\{ \sigma \right\}
\right) \right\} $ is a important one. We hope to keep on a invariant form
of the transformational distribution function $P_{eq}^{\prime }$.

\begin{eqnarray*}
&&R\left\{ P_{{\normalsize eq}}\left( k,\left\{ \sigma \right\} \right)
\right\} \\
&=&\frac 1Z\int_{-\infty }^\infty d\sigma _1d\sigma _2\ldots d\sigma
_N\prod_\beta \delta \left( \mu _1^\beta -\sigma _1^\beta \right) \delta
\left( \mu _2^\beta -\sigma _5^\beta \right) \exp \left[ k\sum_{\left\langle
i,j\right\rangle }\sigma _i\sigma _j-\frac b2\sum_i\sigma _i^2\right] \\
&=&\frac 1Z\prod_\beta \int_{-\infty }^\infty d\sigma _2^\beta d\sigma
_3^\beta d\sigma _4^\beta \exp \left\{ k\left( \mu _1^\beta \sigma _2^\beta
+\sigma _2^\beta \sigma _3^\beta +\sigma _3^\beta \sigma _4^\beta +\sigma
_4^\beta \mu _2^\beta \right) \right. \\
&&\left. -\frac b2\left[ \frac 12\left( \mu _1^\beta \right) ^2+\left(
\sigma _2^\beta \right) ^2+\left( \sigma _3^\beta \right) ^2+\left( \sigma
_4^\beta \right) ^2+\frac 12\left( \mu _2^\beta \right) ^2\right] \right\} \\
&=&\frac 1Z\prod_\beta \sqrt{\frac{\left( 2\pi \right) ^3}{b\left(
b^2-2k^2\right) }}\exp \left\{ \frac{k^4}{b\left( b^2-2k^2\right) }\mu
_1^\beta \mu _2^\beta \right. \\
&&.\left. -\frac b2\frac{b^4-4k^2b^2+2k^4}{b^2\left( b^2-2k^2\right) }\left[ 
\frac 12\left( \mu _1^\beta \right) ^2+\frac 12\left( \mu _2^\beta \right)
^2\right] \right\}
\end{eqnarray*}
Obviously, one must to rescale the spins $\mu _1^\alpha ,$ $\mu _2^\alpha $
and interaction $k$ so as to keep the equilibrium distribution function to
be invariant 
\begin{equation}
\mu ^{\prime }=\xi \left( k\right) \mu =\left( \frac{b^4-4k^2b^2+2k^4}{%
b^2\left( b^2-2k^2\right) }\right) ^{1/2}\mu ,  \eqnum{A.4}  \label{A.4}
\end{equation}
\begin{equation}
k^{\prime }=\frac{k^3b}{b^4-4k^2b^2+2k^4}k,  \eqnum{A.5}  \label{A.5}
\end{equation}
then 
\begin{eqnarray}
R\left\{ P_{{\normalsize eq}}\left( k,\left\{ \sigma \right\} \right)
\right\} &=&\frac 1Z\prod_\beta \sqrt{\frac{\left( 2\pi \right) ^3}{b\left(
b^2-2k^2\right) }}\exp \left\{ k^{\prime }\mu _1^{\prime \beta }\mu
_2^{\prime \beta }-\frac b2\left[ \frac 12\left( \mu _1^{\prime \beta
}\right) ^2+\frac 12\left( \mu _2^{\prime \beta }\right) ^2\right] \right\} 
\nonumber \\
&=&\frac 1{Z^{\prime }}\exp \left\{ k^{\prime }\sum_\beta \mu _1^{\prime
\beta }\mu _2^{\prime \beta }-\frac b2\sum_\beta \left[ \frac 12\left( \mu
_1^{\prime \beta }\right) ^2+\frac 12\left( \mu _2^{\prime \beta }\right)
^2\right] \right\}  \nonumber \\
&=&P_{{\normalsize eq}}^{\prime }\left( \left\{ k^{\prime },\mu ^{\prime
}\right\} \right) .  \eqnum{A.6}  \label{A.6}
\end{eqnarray}

Eq. (\ref{A.5}) is reputed to be the recursion relation which enable one to
determine the fixed point of the static RG transformation $k_c$.

Upon that, we have 
\begin{equation}
R\left\{ \sigma _1^\alpha P_{eq}\left( \left\{ \sigma \right\} \right)
\right\} =\mu _1^\alpha P_{{\normalsize eq}}^{\prime }\left( k^{\prime
},\left\{ \mu ^{\prime }\right\} \right) ,  \eqnum{A.7}  \label{A.7}
\end{equation}
\begin{equation}
R\left\{ \sigma _5^\alpha P_{{\normalsize eq}}\left( \left\{ \sigma \right\}
\right) \right\} =\mu _2^\alpha P_{{\normalsize eq}}^{\prime }\left(
k^{\prime },\left\{ \mu ^{\prime }\right\} \right) ,  \eqnum{A.8}
\label{A.8}
\end{equation}
\begin{equation}
R\left\{ \sigma _2^\alpha P_{{\normalsize eq}}\left( k,\left\{ \sigma
\right\} \right) \right\} =\frac kb\left[ \mu _1^\alpha +\frac{k^2}{b^2-2k^2}%
\left( \mu _1^\alpha +\mu _2^\alpha \right) \right] P_{{\normalsize eq}%
}^{\prime }\left( k^{\prime },\left\{ \mu ^{\prime }\right\} \right) , 
\eqnum{A.9}  \label{A.9}
\end{equation}
\begin{equation}
R\left\{ \sigma _3^\alpha P_{{\normalsize eq}}\left( k,\left\{ \sigma
\right\} \right) \right\} =\frac{k^2}{b^2-2k^2}\left( \mu _1^\alpha +\mu
_2^\alpha \right) P_{{\normalsize eq}}^{\prime }\left( k^{\prime },\left\{
\mu ^{\prime }\right\} \right) ,  \eqnum{A.10}  \label{A.10}
\end{equation}
\begin{equation}
R\left\{ \sigma _4^\alpha P_{{\normalsize eq}}\left( k,\left\{ \sigma
\right\} \right) \right\} =\frac kb\left[ \mu _2^\alpha +\frac{k^2}{b^2-2k^2}%
\left( \mu _1^\alpha +\mu _2^\alpha \right) \right] P_{{\normalsize eq}%
}^{\prime }\left( k^{\prime },\left\{ \mu ^{\prime }\right\} \right) . 
\eqnum{A.11}  \label{A.11}
\end{equation}

Substituting Eqs. (\ref{A.7}-\ref{A.11}) into (\ref{A.2}), one can obtain 
\begin{eqnarray}
&&\left( \frac d{dt}\right) \frac 1{\xi \left( k\right) }\frac{b\left(
b+2k\right) }{b^2-2k^2}h\left( t\right) \sum_\alpha \left( \frac 12\mu
_1^{\prime \alpha }+\frac 12\mu _2^{\prime \alpha }\right) P_{{\normalsize eq%
}}^{\prime }\left( k^{\prime },\left\{ \mu ^{\prime }\right\} \right) 
\nonumber \\
&=&-\frac 1{\xi \left( k\right) }\frac{b^2-4k^2}{b^2-2k^2}h\left( t\right)
\sum_\alpha \left( \frac 12\mu _1^{\prime \alpha }+\frac 12\mu _2^{\prime
\alpha }\right) P_{{\normalsize eq}}^{\prime }\left( k^{\prime },\left\{ \mu
^{\prime }\right\} \right) .  \eqnum{A.12}  \label{A.12}
\end{eqnarray}
Obviously, if the summation for $\alpha $ is arranged in next stage of
iteration, the Eq. (\ref{A.12}) can be written as 
\begin{eqnarray}
&&\left( \frac d{dt}\right) \frac 1{\xi \left( k\right) }\frac{b\left(
b+2k\right) }{b^2-2k^2}h\left( t\right) \sum_\beta \left( \frac 12\mu
_1^{\prime \beta }+\mu _2^{\prime \beta }+\mu _3^{\prime \beta }+\mu
_4^{\prime \beta }+\frac 12\mu _5^{\prime \beta }\right) P_{{\normalsize eq}%
}^{\prime }\left( k^{\prime },\left\{ \mu ^{\prime }\right\} \right) 
\nonumber \\
&=&-\frac 1{\xi \left( k\right) }\frac{b^2-4k^2}{b^2-2k^2}h\left( t\right)
\sum_\beta \left( \frac 12\mu _1^{\prime \beta }+\mu _2^{\prime \beta }+\mu
_3^{\prime \beta }+\mu _4^{\prime \beta }+\frac 12\mu _5^{\prime \beta
}\right) P_{{\normalsize eq}}^{\prime }\left( k^{\prime },\left\{ \mu
^{\prime }\right\} \right) .  \eqnum{A.13}  \label{A.13}
\end{eqnarray}
It is just the Eq. (\ref{3.2.7}).

\section{RG calculation of BKC}

\label{app-b}

\subsection{The RG transformation of $\alpha $-th generator is equivalent to 
$\beta $-th}

\label{app-b1}

We can show that the RG transformation of $\alpha $-th generator is
equivalent to $\beta $-th, but the precondition is that the Gaussian type
distribution constants depend on the coordination number and satisfy the
relation (\ref{3.1.3}). It can be realized via the following calculations.

The effective Hamiltonian of the $\beta $-th generator is 
\begin{eqnarray}
-\frac 1{K_BT}{\cal H}_{{\normalsize eff}}^\beta \left( \sigma ,k\right)
&=&k\left( \sigma _1^\beta \sigma _2^\beta +\sigma _2^\beta \sigma _3^\beta
+\sigma _2^\beta \sigma _4^\beta +\sigma _3^\beta \sigma _4^\beta +\sigma
_4^\beta \sigma _5^\beta \right)  \nonumber \\
&&-\frac{b_3}2\left[ \frac 13\left( \sigma _1^\beta \right) ^2+\left( \sigma
_2^\beta \right) ^2+\left( \sigma _4^\beta \right) ^2+\frac 13\left( \sigma
_5^\beta \right) ^2\right] -\frac{b_2}2\left( \sigma _3^\beta \right) ^2 
\eqnum{B.1.1}  \label{B.1.1}
\end{eqnarray}
where, the coefficient $1/3$ in the terms $\left( \sigma _1^\beta \right) ^2$%
and $\left( \sigma _5^\beta \right) ^2$ comes from the fact that three
neighboring generators share the same site $1$ and $5$. We take the
decimation renormalization transformation operator as 
\begin{equation}
T^\beta \left( \mu ,\sigma \right) =\delta \left( \mu _1^\beta -\sigma
_1^\beta \right) \delta \left( \mu _2^\beta -\sigma _5^\beta \right) , 
\eqnum{B.1.2}  \label{B.1.2}
\end{equation}
then, by integrating spins $\sigma _2,$ $\sigma _3$ and $\sigma _4$ from $%
-\infty $ to $+\infty $, one obtain 
\begin{eqnarray}
&&R\left\{ \exp \left[ -\frac 1{K_BT}{\cal H}_{{\normalsize eff}}^\beta
\left( \sigma ,k\right) \right] \right\}  \nonumber \\
&=&\int_{-\infty }^\infty d\sigma _1d\sigma _2\ldots d\sigma _5\cdot T^\beta
\left( \mu ,\sigma \right) \exp \left[ -\frac 1{K_BT}{\cal H}_{{\normalsize %
eff}}^\beta \left( \sigma ,k\right) \right]  \nonumber \\
&=&C\exp \left\{ k_0\mu _1^\beta \mu _2^\beta -\frac{b_3}2\xi \left[ \frac 13%
\left( \mu _1^\beta \right) ^2+\frac 13\left( \mu _2^\beta \right) ^2\right]
\right\} ,  \eqnum{B.1.3}  \label{B.1.3}
\end{eqnarray}
where 
\[
C=\sqrt{\frac{\left( 2\pi \right) ^3}{b_2b_3^2\left(
1-k/b_3-2k^2/b_2b_3\right) }}, 
\]
\[
k_0=\frac{k^3\left( k+b_2\right) }{\left( b_2b_3-kb_2-2k^2\right) \left(
k+b_3\right) }, 
\]
\[
\xi =\frac{3k^4-2k^3b_3-4k^2b_2b_3-2b_3^2k^2+b_2b_3^3}{b_3\left(
b_2b_3-kb_2-2k^2\right) \left( k+b_3\right) }, 
\]
if taking 
\begin{equation}
\mu ^{\prime }=\sqrt{\xi }\mu =\sqrt{\frac{%
3k^4-2k^3b_3-4k^2b_2b_3-2b_3^2k^2+b_2b_3^3}{b_3\left(
b_2b_3-kb_2-2k^2\right) \left( k+b_3\right) }}\mu ,  \eqnum{B.1.4}
\label{B.1.4}
\end{equation}
\begin{equation}
k^{\prime }=\frac{k_0}\xi =\frac{k^3\left( k+b_2\right) b_3}{%
3k^4-2b_3k^3-4k^2b_2b_3-2k^2b_3^2+b_2b_3^3},  \eqnum{B.1.5}  \label{B.1.5}
\end{equation}
then 
\begin{equation}
R\left\{ \exp \left[ -\frac 1{K_BT}{\cal H}_{{\normalsize eff}}^\beta \left(
\sigma ,k\right) \right] \right\} =C\exp \left\{ k^{\prime }\mu _1^{\prime
\beta }\mu _2^{\prime \beta }-\frac{b_3}2\left[ \frac 13\left( \mu
_1^{\prime \beta }\right) ^2+\frac 13\left( \mu _2^{\prime \beta }\right)
^2\right] \right\} .  \eqnum{B.1.6}  \label{B.1.6}
\end{equation}

The same as above, the effective Hamiltonian of the $\alpha $-th generator
is 
\begin{eqnarray}
-\frac 1{K_BT}{\cal H}_{{\normalsize eff}}^\alpha \left( \sigma ,k\right)
&=&k\left( \sigma _1^\alpha \sigma _2^\alpha +\sigma _2^\alpha \sigma
_3^\alpha +\sigma _2^\alpha \sigma _4^\alpha +\sigma _3^\alpha \sigma
_4^\alpha +\sigma _4^\alpha \sigma _5^\alpha \right)  \nonumber \\
&&-\frac{b_3}2\left[ \frac 13\left( \sigma _1^\alpha \right) ^2+\left(
\sigma _2^\alpha \right) ^2+\left( \sigma _4^\alpha \right) ^2\right] -\frac{%
b_2}2\left[ \left( \sigma _3^\alpha \right) ^2+\frac 12\left( \sigma
_5^\alpha \right) ^2\right] ,  \eqnum{B.1.7}  \label{B.1.7}
\end{eqnarray}
where, the coefficient $1/3$ (or $1/2$) in the terms $\left( \sigma
_1^\alpha \right) ^2$ (or $\left( \sigma _5^\alpha \right) ^2$ ) comes from
the fact that three (or two) neighboring generators share the same site $1$
(or $5$ ). By integrating spins $\sigma _2,$ $\sigma _3$ and $\sigma _4$
from $-\infty $ to $+\infty $, one obtain 
\begin{eqnarray}
&&R\left\{ \exp \left[ -\frac 1{K_BT}{\cal H}_{{\normalsize eff}}^\alpha
\left( \sigma ,k\right) \right] \right\}  \nonumber \\
&=&\int_{-\infty }^\infty d\sigma _1d\sigma _2\ldots d\sigma _5\cdot
T^\alpha \left( \mu ,\sigma \right) \exp \left[ -\frac 1{K_BT}{\cal H}_{%
{\normalsize eff}}^\alpha \left( \sigma ,k\right) \right]  \nonumber \\
&=&C\exp \left\{ k_0\mu _1^\alpha \mu _2^\alpha -\frac{b_3}2\xi _1\frac 13%
\left( \mu _1^\alpha \right) ^2-\frac{b_2}2\xi _2\frac 12\left( \mu
_2^\alpha \right) ^2\right\} ,  \eqnum{B.1.8}  \label{B.1.8}
\end{eqnarray}
where 
\[
k_0=\frac{k^3\left( k+b_2\right) }{\left( b_2b_3-kb_2-2k^2\right) \left(
k+b_3\right) }, 
\]
\begin{eqnarray*}
\xi _1 &=&1-3k^2/b_3^2-\frac{3k^4/b_3^4}{1-k^2/b_3^2}-\frac{3k^4/b_2b_3^3}{%
\left( 1-k/b_3\right) \left( 1-k/b_3-2k^2/b_2b_3\right) }, \\
\xi _2 &=&1-\frac{2k^2/b_2b_3}{1-k^2/b_3^2}-\frac{2k^4/b_2^2b_3^2}{\left(
1-k/b_3-2k^2/b_2b_3\right) \left( 1-k/b_3\right) }.
\end{eqnarray*}
Based on the relation (\ref{3.1.3}), i.e., 
\[
b_2/b_3=2/3, 
\]
we can see 
\[
\xi _1=\xi _2=\xi =\frac{8k^4-8k^3b_2-28b_2^2k^2+9b_2^4}{b_2\left(
3b_2^2-2kb_2-4k^2\right) \left( 2k+3b_2\right) }. 
\]
So, if only 
\begin{equation}
\mu ^{\prime }=\sqrt{\xi }\mu =\sqrt{\frac{8k^4-8k^3b_2-28b_2^2k^2+9b_2^4}{%
b_2\left( 3b_2^2-2kb_2-4k^2\right) \left( 2k+3b_2\right) }}\mu , 
\eqnum{B.1.9}  \label{B.1.9}
\end{equation}
\begin{equation}
k^{\prime }=\frac{k_0}\xi =\frac{4b_2k^3\left( k+b_2\right) }{%
8k^4-8k^3b_2-28b_2^2k^2+9b_2^4},  \eqnum{B.1.10}  \label{B.1.10}
\end{equation}
one can get 
\begin{equation}
R\left\{ -\frac 1{K_BT}{\cal H}_{{\normalsize eff}}^\alpha \left( \sigma
,k\right) \right\} =C\exp \left\{ k^{\prime }\mu _1^{\prime \alpha }\mu
_2^{\prime \alpha }-\frac{b_3}2\frac 13\left( \mu _1^{\prime \alpha }\right)
^2-\frac{b_2}2\frac 12\left( \mu _2^{\prime \alpha }\right) ^2\right\} . 
\eqnum{B.1.11}  \label{B.1.11}
\end{equation}
In fact, (\ref{B.1.4}-\ref{B.1.5}) is coincided with (\ref{B.1.9}-\ref
{B.1.10}). Solving fixed-point equation $k^{*}=k^{\prime }=k,$ the critical
point $k_c$ is obtained 
\[
k_c=\frac{b_2}2=\frac{b_3}3. 
\]

Now, as the decimation renormalization transformation operator is taken as 
\[
T\left( \mu ,\sigma \right) =\prod_\alpha \delta \left( \mu _1^\alpha
-\sigma _1^\alpha \right) \delta \left( \mu _2^\alpha -\sigma _5^\alpha
\right) , 
\]
from (\ref{B.1.6}) and (\ref{B.1.11}) we can see 
\begin{equation}
R\left\{ P_{{\normalsize eq}}\left( k,\left\{ \sigma \right\} \right)
\right\} =\frac 1{Z^{\prime }}\exp \left[ k^{\prime }\sum_{\left\langle
i,j\right\rangle }\mu _i^{\prime }\mu _j^{\prime }-\frac{b_{q_i}}2\sum_i\mu
_i^{\prime 2}\right] =P_{{\normalsize eq}}^{\prime }\left( k^{\prime
},\left\{ \mu ^{\prime }\right\} \right)  \eqnum{B.1.12}  \label{B.1.12}
\end{equation}
This means that the distribution function is invariant under RG
transformation.

\subsection{RG transformation of master equation}

\label{app-b2}

Besides $R\left\{ P_{eq}\left( k,\left\{ \sigma \right\} \right) \right\} $,
one can also calculate 
\begin{eqnarray}
R\left\{ \sigma _i^\alpha P_{{\normalsize eq}}\left( k,\left\{ \sigma
\right\} \right) \right\} &=&\sum_{\left\{ \sigma \right\} }\left\{ T\left(
\mu ,\sigma \right) \sigma _i^\alpha P_{{\normalsize eq}}\left( k,\left\{
\sigma \right\} \right) \right\}  \nonumber \\
&=&P_{{\normalsize eq}}^{\prime }\left( k^{\prime },\left\{ \mu ^{\prime
}\right\} \right) \cdot \frac{W_{\sigma _i^\alpha }}W  \eqnum{B.2.1}
\label{B.2.1}
\end{eqnarray}
where 
\begin{eqnarray}
W &=&\int_{-\infty }^\infty d\sigma _2^\alpha d\sigma _3^\alpha d\sigma
_4^\alpha \exp \left\{ k\left( \mu _1^\alpha \sigma _2^\alpha +\sigma
_2^\alpha \sigma _3^\alpha +\sigma _2^\alpha \sigma _4^\alpha +\sigma
_3^\alpha \sigma _4^\alpha +\sigma _4^\alpha \mu _2^\alpha \right) \right. 
\nonumber \\
&&\left. -\frac{b_3}2\left[ \left( \sigma _2^\alpha \right) ^2+\left( \sigma
_4^\alpha \right) ^2\right] -\frac{b_2}2\left( \sigma _3^\alpha \right)
^2\right\}  \nonumber \\
&=&\sqrt{\frac{\left( 2\pi \right) ^3}{b_3\left( b_2b_3-kb_2-2k^2\right) }}%
\exp \left\{ \left( \frac{k^3\left( b_2+k\right) }{\left(
b_2b_3-kb_2-2k^2\right) \left( b_3+k\right) }\mu _1^\alpha \mu _2^\alpha
\right. \right.  \nonumber \\
&&\left. \left. +\frac{k^2\left( b_2b_3-k^2\right) }{\left(
b_2b_3-kb_2-2k^2\right) \left( b_3+k\right) }\left[ \frac 12\left( \mu
_1^\alpha \right) ^2+\frac 12\left( \mu _2^\alpha \right) ^2\right] \right)
\right\} ,  \eqnum{B.2.2}  \label{B.2.2}
\end{eqnarray}
\begin{eqnarray}
W_{\sigma _i^\alpha } &=&\int_{-\infty }^\infty d\sigma _2^\alpha d\sigma
_3^\alpha d\sigma _4^\alpha \cdot \sigma _i^\alpha \exp \left\{ k\left( \mu
_1^\alpha \sigma _2^\alpha +\sigma _2^\alpha \sigma _3^\alpha +\sigma
_2^\alpha \sigma _4^\alpha +\sigma _3^\alpha \sigma _4^\alpha +\sigma
_4^\alpha \mu _2^\alpha \right) \right.  \nonumber \\
&&\left. -\frac{b_3}2\left[ \left( \sigma _2^\alpha \right) ^2+\left( \sigma
_4^\alpha \right) ^2\right] -\frac{b_2}2\left( \sigma _3^\alpha \right)
^2\right\} ,  \eqnum{B.2.3}  \label{B.2.3}
\end{eqnarray}
Upon that, we have 
\begin{equation}
R\left\{ \sigma _1^\alpha P_{{\normalsize eq}}\left( \left\{ k,\sigma
\right\} \right) \right\} =\frac 1{\sqrt{\xi }}\mu _1^{\prime \alpha }P_{%
{\normalsize eq}}^{\prime }\left( \left\{ k^{\prime },\mu ^{\prime }\right\}
\right) ,  \eqnum{B.2.4}  \label{B.2.4}
\end{equation}
\begin{equation}
R\left\{ \sigma _5^\alpha P_{{\normalsize eq}}\left( \left\{ k,\sigma
\right\} \right) \right\} =\frac 1{\sqrt{\xi }}\mu _2^{\prime \alpha }P_{%
{\normalsize eq}}^{\prime }\left( \left\{ k^{\prime },\mu ^{\prime }\right\}
\right) ,  \eqnum{B.2.5}  \label{B.2.5}
\end{equation}
\begin{equation}
R\left\{ \sigma _2^\alpha P_{{\normalsize eq}}\left( \left\{ k,\sigma
\right\} \right) \right\} =\frac 1{\sqrt{\xi }}\frac{k\left(
b_3b_2-k^2\right) \mu _1^{\prime \alpha }+k^2\left( k+b_2\right) \mu
_2^{\prime \alpha }}{\left( k+b_3\right) \left( b_2b_3-kb_2-2k^2\right) }P_{%
{\normalsize eq}}^{\prime }\left( \left\{ k^{\prime },\mu ^{\prime }\right\}
\right) ,  \eqnum{B.2.6}  \label{B.2.6}
\end{equation}
\begin{equation}
R\left\{ \sigma _3^\alpha P_{{\normalsize eq}}\left( \left\{ k,\sigma
\right\} \right) \right\} =\frac 1{\sqrt{\xi }}\frac{k^2}{b_2b_3-kb_2-2k^2}%
\left( \mu _1^{\prime \alpha }+\mu _2^{\prime \alpha }\right) P_{%
{\normalsize eq}}^{\prime }\left( \left\{ k^{\prime },\mu ^{\prime }\right\}
\right) ,  \eqnum{B.2.7}  \label{B.2.7}
\end{equation}
\begin{equation}
R\left\{ \sigma _4^\alpha P_{{\normalsize eq}}\left( \left\{ k,\sigma
\right\} \right) \right\} =\frac 1{\sqrt{\xi }}\frac{k^2\left( k+b_2\right)
\mu _1^{\prime \alpha }+k\left( b_3b_2-k^2\right) \mu _2^{\prime \alpha }}{%
\left( k+b_3\right) \left( b_2b_3-kb_2-2k^2\right) }P_{{\normalsize eq}%
}^{\prime }\left( \left\{ k^{\prime },\mu ^{\prime }\right\} \right) . 
\eqnum{B.2.8}  \label{B.2.8}
\end{equation}

By virtue of these integral results, Eqs. (\ref{3.3.4}) and (\ref{3.3.5})
can be obtained.

\section{RG calculation of MBKC}

\label{app-c}

\subsection{The RG transformation of $\alpha $-th generator is equivalent to 
$\beta $-th}

\label{app-c1}

For MBKC, we can also show that the RG transformation of $\alpha $-th
generator is equivalent to $\beta $-th, but the precondition is, the same as
about BKC, that the Gaussian type distribution constants depend on the
coordination number and satisfy the relation (\ref{3.1.3}). It can be
realized via the following calculations.

The effective Hamiltonian of the $\alpha $-th generator is 
\begin{eqnarray}
&&-\frac 1{K_BT}{\cal H}_{{\normalsize eff}}^\alpha \left( \sigma ^\alpha
,k\right)  \nonumber \\
&=&k\left[ \sigma _1^\alpha \sigma _3^\alpha +\sigma _3^\alpha \sigma
_4^\alpha +\sigma _4^\alpha \sigma _2^\alpha +\sum_{i=5}^{m+4}\left( \sigma
_4^\alpha +\sigma _3^\alpha \right) \sigma _i^\alpha \right]  \nonumber \\
&&-\frac{b_{m+2}}2\left[ \frac 1{m+2}\left( \sigma _1^\alpha \right)
^2+\left( \sigma _3^\alpha \right) ^2+\left( \sigma _4^\alpha \right)
^2\right] -\frac{b_2}2\left[ \sum_{i=5}^{m+4}\left( \sigma _i^\alpha \right)
^2+\frac 12\left( \sigma _2^\alpha \right) ^2\right] ,  \eqnum{C.1.1}
\label{C.1.1}
\end{eqnarray}
where, the coefficient $1/\left( m+2\right) $ (or $1/2$) in the terms $%
\left( \sigma _1^\alpha \right) ^2$ (or $\left( \sigma _2^\alpha \right) ^2$
) comes from the fact that $\left( m+2\right) $ (or $2$ ) neighboring
generators share the same site $1$ (or $2$ ). We take the decimation
renormalization transformation operator as 
\begin{equation}
T^\alpha \left( \mu ,\sigma ^\alpha \right) =\delta \left( \mu _1^\alpha
-\sigma _1^\alpha \right) \delta \left( \mu _2^\alpha -\sigma _2^\alpha
\right) ,  \eqnum{C.1.2}  \label{C.1.2}
\end{equation}
then, by integrating spins $\sigma _3,$ $\sigma _4$ and $\sigma _i$ $%
(i=5,\cdots ,m+4)$ from $-\infty $ to $+\infty $, one obtain 
\begin{eqnarray}
&&R\left\{ \exp \left[ -\frac 1{K_BT}{\cal H}_{{\normalsize eff}}^\alpha
\left( \sigma ^\alpha ,k\right) \right] \right\}  \nonumber \\
&=&\int_{-\infty }^\infty \prod_{i=1}^{m+4}d\sigma _i^\alpha \cdot T^\alpha
\left( \mu ^\alpha ,\sigma ^\alpha \right) \exp \left[ -\frac 1{K_BT}{\cal H}%
_{{\normalsize eff}}^\alpha \left( \sigma ^\alpha ,k\right) \right] 
\nonumber \\
&=&C\exp \left\{ k_0\mu _1^\alpha \mu _2^\alpha -\frac{b_{m+2}}2\frac 1{m+2}%
\xi _1\left( \mu _1^\alpha \right) ^2-\frac{b_2}2\frac 12\xi _2\left( \mu
_2^\alpha \right) ^2\right\}  \eqnum{C.1.3}  \label{C.1.3}
\end{eqnarray}
where 
\[
C=\left( \frac{2\pi }{b_2}\right) ^{m/2}\sqrt{\frac{\left( 2\pi \right) ^2b_2%
}{\left( b_{m+2}+k\right) \left( b_2b_{m+2}-b_2k-2mk^2\right) }}, 
\]
\[
k_0=\frac{k^3\left( b_2+mk\right) }{\left( b_{m+2}+k\right) \left(
b_2b_{m+2}-b_2k-2mk^2\right) }, 
\]
\[
\xi _1=\frac{m\left( m+2\right) k^4-2b_{m+2}mk^3-2b_{m+2}^2mk^2-b_2k^2\left(
m+2\right) b_{m+2}-k^2b_2b_{m+2}+b_{m+2}^3b_2}{b_{m+2}\left(
b_{m+2}+k\right) \left( b_2b_{m+2}-b_2k-2mk^2\right) }, 
\]
\[
\xi _2=\frac{%
-2k^2b_2b_{m+2}+2k^4m-2b_2b_{m+2}mk^2+b_{m+2}^2b_2^2-2b_2mk^3-k^2b_2^2}{%
b_2\left( b_{m+2}+k\right) \left( b_2b_{m+2}-b_2k-2mk^2\right) }. 
\]
Base on the relation (\ref{3.1.3}), i.e., 
\begin{equation}
\frac{b_{m+2}}{b_2}=\frac{m+2}2  \eqnum{C.1.4}  \label{C.1.4}
\end{equation}
we can see 
\begin{eqnarray}
\xi &=&\xi _1=\xi _2  \nonumber \\
&=&\frac{-8k^4m+8b_2mk^3+4\left( m+2\right) b_2^2mk^2+4k^2b_2^2\left(
m+2\right) +4k^2b_2^2-\left( m+2\right) ^2b_2^4}{b_2\left( \left( m+2\right)
b_2+2k\right) \left( -b_2^2\left( m+2\right) +2b_2k+4mk^2\right) }, 
\eqnum{C.1.5}  \label{C.1.5}
\end{eqnarray}
\begin{equation}
k_0=\frac{4k^3\left( b_2+mk\right) }{\left( \left( m+2\right) b_2+2k\right)
\left( b_2^2\left( m+2\right) -2b_2k-4mk^2\right) },  \eqnum{C.1.6}
\label{C.1.6}
\end{equation}
if taking 
\begin{equation}
\mu ^{\prime }=\sqrt{\xi }\mu ,k^{\prime }=\frac{k_0}\xi ,  \eqnum{C.1.7}
\label{C.1.7}
\end{equation}
then 
\begin{equation}
R\left\{ \exp \left[ -\frac 1{K_BT}{\cal H}_{{\normalsize eff}}^\alpha
\left( \sigma ,k\right) \right] \right\} =C\exp \left\{ k^{\prime }\mu
_1^{\prime \alpha }\mu _2^{\prime \alpha }-\frac{b_{m+2}}2\frac 1{m+2}\left(
\mu _1^{\prime \alpha }\right) ^2-\frac{b_2}2\frac 12\left( \mu _5^{\prime
\alpha }\right) ^2\right\} .  \eqnum{C.1.8}  \label{C.1.8}
\end{equation}

The same as above for case (2), the effective Hamiltonian of the $\beta $-th
generator is 
\begin{eqnarray}
&&-\frac 1{K_BT}{\cal H}_{{\normalsize eff}}^\beta \left( \sigma ,k\right) 
\nonumber \\
&=&k\left[ \sigma _1^\beta \sigma _3^\beta +\sigma _4^\beta \sigma _2^\beta
+\sum_{i=5}^{m+4}\left( \sigma _3^\beta +\sigma _3^\beta \right) \sigma
_i^\beta \right]  \nonumber \\
&&-\frac{b_{m+2}}2\left[ \frac 1{m+2}\left( \sigma _1^\beta \right)
^2+\left( \sigma _3^\beta \right) ^2+\left( \sigma _4^\beta \right) ^2+\frac %
1{m+2}\left( \sigma _2^\beta \right) ^2\right] -\frac{b_2}2\left[
\sum_{i=5}^{m+4}\left( \sigma _i^\beta \right) ^2\right] ,  \eqnum{C.1.9}
\label{C.1.9}
\end{eqnarray}
where the coefficients $1/\left( m+2\right) $ in the terms $\left( \sigma
_1^\beta \right) ^2$ and $\left( \sigma _2^\beta \right) ^2$ come from the
fact that $m+2$ neighboring generators share the same site $1$ and $5$. By
integrating spins $\sigma _3,$ $\sigma _4$ and $\sigma _i$ $(i=5,\cdots
,m+4) $ from $-\infty $ to $+\infty $ and using the relation (\ref{C.1.4}),
the same result can be obtain 
\begin{eqnarray}
&&R\left\{ \exp \left[ -\frac 1{K_BT}{\cal H}_{{\normalsize eff}}^\beta
\left( \sigma ^\beta ,k\right) \right] \right\}  \nonumber \\
&=&\int_{-\infty }^\infty \prod_{i=1}^{m+4}d\sigma _i^\beta \cdot T^\beta
\left( \mu ,\sigma ^\beta \right) \exp \left[ -\frac 1{K_BT}{\cal H}_{%
{\normalsize eff}}^\beta \left( \sigma ^\beta ,k\right) \right]  \nonumber \\
&=&C\exp \left\{ k^{\prime }\mu _1^{\prime \beta }\mu _2^{\prime \beta }-%
\frac{b_{m+2}}2\left[ \frac 1{m+2}\left( \mu _1^{\prime \beta }\right) ^2+%
\frac 1{m+2}\left( \mu _2^{\prime \beta }\right) ^2\right] \right\} . 
\eqnum{C.1.10}  \label{C.1.10}
\end{eqnarray}

Therefore, from (\ref{C.1.8}) and (\ref{C.1.10}), we have 
\begin{equation}
R\left\{ P_{{\normalsize eq}}\left( k,\left\{ \sigma \right\} \right)
\right\} =\frac 1{Z^{\prime }}\exp \left[ k^{\prime }\sum_{\left\langle
i,j\right\rangle }\mu _i^{\prime }\mu _j^{\prime }-\frac{b_{q_i}}2\sum_i\mu
_i^{\prime 2}\right] =P_{{\normalsize eq}}^{\prime }\left( k^{\prime
},\left\{ \mu ^{\prime }\right\} \right) ,  \eqnum{C.1.11}  \label{C.1.11}
\end{equation}
where the recursion relation is 
\begin{equation}
k^{\prime }=\frac{4k^2\left( b_2+mk\right) b_2}{8k^4m-8b_2mk^3-4\left(
m^2+3m+3\right) b_2^2k^2+\left( m+2\right) ^2b_2^4}k.  \eqnum{C.1.12}
\label{C.1.12}
\end{equation}
$\allowbreak $Expression (\ref{C.1.11}) means that the distribution function
is invariant under RG transformation.

\subsection{RG transformation of master equation}

\label{app-c2}

Besides $R\left\{ P_{eq}\left( k,\left\{ \sigma \right\} \right) \right\} $,
one can also calculate 
\begin{eqnarray*}
R\left\{ \sigma _j^\alpha P_{{\normalsize eq}}\left( k,\left\{ \sigma
\right\} \right) \right\} &=&\sum_{\left\{ \sigma \right\} }\left\{ T\left(
\mu ,\sigma \right) \sigma _j^\alpha P_{{\normalsize eq}}\left( k,\left\{
\sigma \right\} \right) \right\} \\
&=&\int_{-\infty }^\infty d\sigma _1d\sigma _2\ldots d\sigma _N\prod_\beta
\delta \left( \mu _1^\alpha -\sigma _1^\alpha \right) \delta \left( \mu
_2^\alpha -\sigma _2^\alpha \right) \sigma _j^\alpha P_{{\normalsize eq}%
}\left( k,\left\{ \sigma \right\} \right) \\
&=&P_{{\normalsize eq}}^{\prime }\left( k^{\prime },\left\{ \mu ^{\prime
}\right\} \right) \cdot \frac{W_{\sigma _j^\alpha }}W,
\end{eqnarray*}
where 
\begin{eqnarray*}
W &=&\int_{-\infty }^\infty d\sigma _3^\alpha d\sigma _4^\alpha \left(
\prod_{i=5}^{m+4}d\sigma _i^\alpha \right) \exp \left\{ k\left[ \mu
_1^\alpha \sigma _3^\alpha +\sigma _3^\alpha \sigma _4^\alpha +\sigma
_4^\alpha \mu _2^\alpha +\sum_{i=5}^{m+4}\left( \sigma _4^\alpha +\sigma
_3^\alpha \right) \sigma _i^\alpha \right] \right. \\
&&\left. -\frac{b_{m+1}}2\left[ \left( \sigma _3^\alpha \right) ^2+\left(
\sigma _4^\alpha \right) ^2\right] -\frac{b_2}2\left[ \sum_{i=5}^{m+4}\left(
\sigma _i^\alpha \right) ^2\right] \right\} \\
&=&\left( \frac{2\pi }{b_2}\right) ^{m/2}\sqrt{\frac{\left( 2\pi \right)
^2b_2}{\left( b_{m+2}+k\right) \left( b_2b_{m+2}-b_2k-2mk^2\right) }}\exp
\left\{ \frac{k^3\left( b_2+mk\right) }{\left( b_{m+2}+k\right) \left(
b_2b_{m+2}-b_2k-2mk^2\right) }\mu _1^\alpha \mu _2^\alpha \right. \\
&&\left. +\frac 12\frac{k^2\left( b_2b_{m+2}-mk^2\right) }{\left(
b_{m+2}+k\right) \left( b_2b_{m+2}-b_2k-2mk^2\right) }\left[ \left( \mu
_1^\alpha \right) ^2+\left( \mu _2^\alpha \right) ^2\right] \right\} ,
\end{eqnarray*}
\begin{eqnarray*}
&&W_{\sigma _j^\alpha }\stackrel{j=5,\ldots ,m+4}{=}\int_{-\infty }^\infty
d\sigma _3^\alpha d\sigma _4^\alpha \left( \prod_{i=5}^{m+4}d\sigma
_i^\alpha \right) \sigma _j^\alpha \exp \left\{ k\left[ \mu _1^\alpha \sigma
_3^\alpha +\sigma _3^\alpha \sigma _4^\alpha +\sigma _4^\alpha \mu _2^\alpha
+\sum_{i=5}^{m+4}\left( \sigma _4^\alpha +\sigma _3^\alpha \right) \sigma
_i^\alpha \right] \right. \\
&&\left. -\frac{b_{m+2}}2\left[ \left( \sigma _3^\alpha \right) ^2+\left(
\sigma _4^\alpha \right) ^2\right] -\frac{b_2}2\left[ \sum_{i=5}^{m+4}\left(
\sigma _i^\alpha \right) ^2\right] \right\} \\
&=&\int_{-\infty }^\infty d\sigma _3^\alpha d\sigma _4^\alpha \left(
\prod_{i\neq j=5}^{m+4}\left( \int_{-\infty }^\infty d\sigma _i^\alpha \exp
\left[ k\left( \sigma _4^\alpha +\sigma _3^\alpha \right) \sigma _i^\alpha -%
\frac{b_2}2\left( \sigma _i^\alpha \right) ^2\right] \right) \right) \\
&&\times \left( \int_{-\infty }^\infty d\sigma _j^\alpha \cdot \sigma
_j^\alpha \exp \left[ k\left( \sigma _4^\alpha +\sigma _3^\alpha \right)
\sigma _j^\alpha -\frac{b_2}2\left( \sigma _j^\alpha \right) ^2\right]
\right) \\
&&\times \exp \left\{ k\left[ \mu _1^\alpha \sigma _3^\alpha +\sigma
_3^\alpha \sigma _4^\alpha +\sigma _4^\alpha \mu _2^\alpha \right] -\frac{%
b_{m+2}}2\left[ \left( \sigma _3^\alpha \right) ^2+\left( \sigma _4^\alpha
\right) ^2\right] \right\} \\
&=&\frac{k^2}{b_2b_{m+2}-b_2k-2mk^2}W,
\end{eqnarray*}
\begin{eqnarray*}
W_{\sigma _4^\alpha } &=&\int_{-\infty }^\infty d\sigma _3^\alpha d\sigma
_4^\alpha \left( \prod_{i=5}^{m+4}d\sigma _i^\alpha \right) \sigma _4^\alpha
\exp \left\{ k\left[ \mu _1^\alpha \sigma _3^\alpha +\sigma _3^\alpha \sigma
_4^\alpha +\sigma _4^\alpha \mu _2^\alpha +\sum_{i=5}^{m+4}\left( \sigma
_4^\alpha +\sigma _3^\alpha \right) \sigma _i^\alpha \right] \right. \\
&&\left. -\frac{b_{m+2}}2\left[ \left( \sigma _3^\alpha \right) ^2+\left(
\sigma _4^\alpha \right) ^2\right] -\frac{b_2}2\left[ \sum_{i=5}^{m+4}\left(
\sigma _i^\alpha \right) ^2\right] \right\} \\
&=&\frac{\left( mk^3+b_2k^2\right) \mu _1^\alpha +\left(
b_2b_{m+2}-mk^2\right) k\mu _2^\alpha }{\left( k+b_{m+2}\right) \left(
b_2b_{m+2}-b_2k-2mk^2\right) }W,
\end{eqnarray*}
\begin{eqnarray*}
W_{\sigma _3^\alpha } &=&\int_{-\infty }^\infty d\sigma _3^\alpha d\sigma
_4^\alpha \left( \prod_{i=5}^{m+4}d\sigma _i^\alpha \right) \sigma _3^\alpha
\exp \left\{ k\left[ \mu _1^\alpha \sigma _3^\alpha +\sigma _3^\alpha \sigma
_4^\alpha +\sigma _4^\alpha \mu _2^\alpha +\sum_{i=5}^{m+4}\left( \sigma
_4^\alpha +\sigma _3^\alpha \right) \sigma _i^\alpha \right] \right. \\
&&\left. -\frac{b_{m+2}}2\left[ \left( \sigma _3^\alpha \right) ^2+\left(
\sigma _4^\alpha \right) ^2\right] -\frac{b_2}2\left[ \sum_{i=5}^{m+4}\left(
\sigma _i^\alpha \right) ^2\right] \right\} \\
&=&\frac{\left( \left( mk^2+b_2k\right) \mu _2^\alpha +\left(
b_2b_{m+2}-mk^2\right) \mu _1^\alpha \right) k}{\left( k+b_{m+2}\right)
\left( b_2b_{m+2}-b_2k-2mk^2\right) }W,
\end{eqnarray*}
and then 
\begin{equation}
R\left\{ \sigma _1^\alpha P_{{\normalsize eq}}\left( k,\left\{ \sigma
\right\} \right) \right\} =\frac 1{\sqrt{\xi }}\mu _1^{\prime \alpha }P_{%
{\normalsize eq}}^{\prime }\left( k^{\prime },\left\{ \mu ^{\prime }\right\}
\right) ,  \eqnum{C.2.1}  \label{C.2.1}
\end{equation}
\begin{equation}
R\left\{ \sigma _2^\alpha P_{{\normalsize eq}}\left( k,\left\{ \sigma
\right\} \right) \right\} =\frac 1{\sqrt{\xi }}\mu _2^{\prime \alpha }P_{%
{\normalsize eq}}^{\prime }\left( k^{\prime },\left\{ \mu ^{\prime }\right\}
\right) ,  \eqnum{C.2.2}  \label{C.2.2}
\end{equation}
\begin{equation}
R\left\{ \sigma _3^\alpha P_{{\normalsize eq}}\left( k,\left\{ \sigma
\right\} \right) \right\} =\frac 1{\sqrt{\xi }}\frac{\left(
mk^3+b_2k^2\right) \mu _2^{\prime \alpha }+k\left( b_2b_{m+2}-mk^2\right)
\mu _1^{\prime \alpha }}{\left( k+b_{m+2}\right) \left(
b_2b_{m+2}-b_2k-2mk^2\right) }P_{{\normalsize eq}}^{\prime }\left( k^{\prime
},\left\{ \mu ^{\prime }\right\} \right) ,  \eqnum{C.2.3}  \label{C.2.3}
\end{equation}
\begin{equation}
R\left\{ \sigma _4^\alpha P_{{\normalsize eq}}\left( k,\left\{ \sigma
\right\} \right) \right\} =\frac 1{\sqrt{\xi }}\frac{\left(
mk^3+b_2k^2\right) \mu _1^{\prime \alpha }+\left( b_2b_{m+2}-mk^2\right)
k\mu _2^{\prime \alpha }}{\left( k+b_{m+2}\right) \left(
b_2b_{m+2}-b_2k-2mk^2\right) }P_{{\normalsize eq}}^{\prime }\left( k^{\prime
},\left\{ \mu ^{\prime }\right\} \right) ,  \eqnum{C.2.4}  \label{C.2.4}
\end{equation}
\begin{equation}
R\left\{ \sigma _j^\alpha P_{{\normalsize eq}}\left( k,\left\{ \sigma
\right\} \right) \right\} \stackrel{j=5,\ldots ,m+4}{=}\frac 1{\sqrt{\xi }}%
\frac{k^2}{b_2b_{m+2}-b_2k-2mk^2}\left( \mu _1^{\prime \alpha }+\mu
_2^{\prime \alpha }\right) P_{{\normalsize eq}}^{\prime }\left( k^{\prime
},\left\{ \mu ^{\prime }\right\} \right) .  \eqnum{C.2.5}  \label{C.2.5}
\end{equation}
By virtue of these integral results, Eqs. (\ref{3.4.5}) and (\ref{3.4.6})
can be obtained.

\begin{figure}[tbp]
\caption{The second construction stage of the NBKC and the BKC. (a) NBKC:
All of the generations are the same; (b)BKC: there are two kinds of typical
generators (such as $\alpha $-th and $\beta $-th).}
\label{fig2}
\end{figure}

\begin{figure}[tbp]
\caption{Decimation RG procedure :(a) NBKC; (b) BKC.}
\label{fig3}
\end{figure}

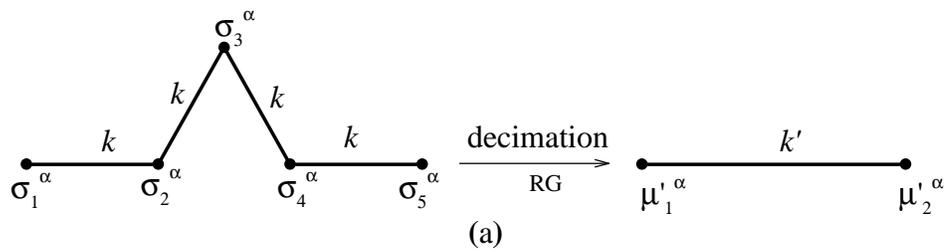
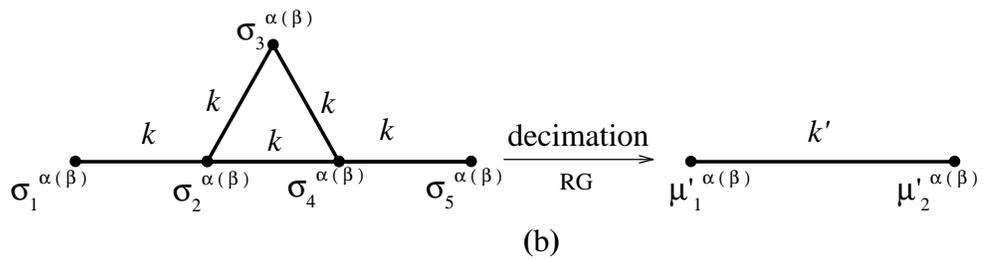
\begin{figure}[tbp]
\caption{The construction procedure of multi-branching Koch curve (MBKC)
with $D_f=\ln (2m+3)/\ln 3$.}
\label{fig4}
\end{figure}

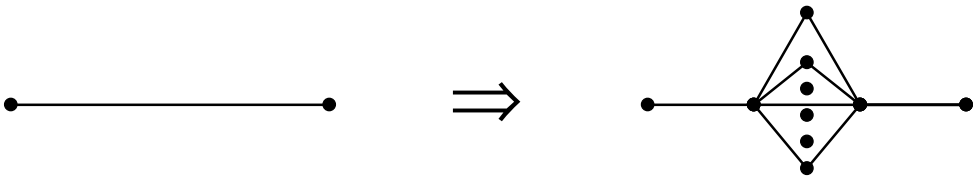
\begin{figure}[tbp]
\caption{Decimation RG procedure of the multi-branching Koch curve (MBKC).}
\label{fig5}
\end{figure}

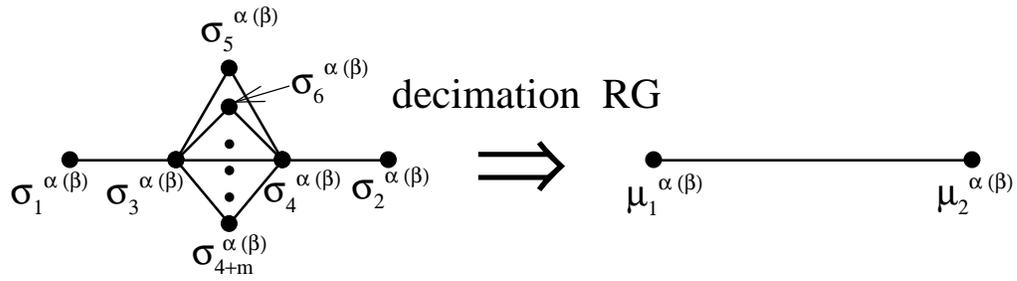
\begin{figure}[tbp]
\caption{Critical dynamical exponent of the kinetic Gaussian model on the
Multi-branching Koch curve.}
\label{fig6}
\end{figure}

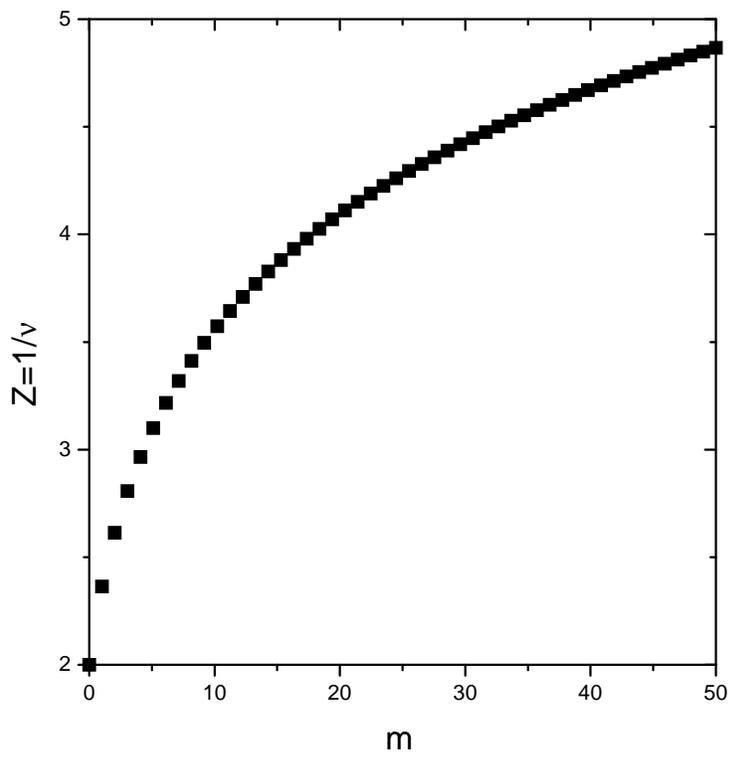
\end{document}